\documentclass[preprint,a4paper, 11pt]{aastex631}
\usepackage[varg]{txfonts}%
\usepackage{graphicx, natbib}%
\usepackage{enumerate}%
\usepackage{etoolbox}%
\bibpunct{(}{)}{,}{a}{}{,}%
\usepackage[pagewise,mathlines]{lineno}%
\usepackage{textcomp}
\usepackage{gensymb}
\usepackage{enumitem}
\usepackage{hyperref}
\setlist{nosep} % or \setlist{noitemsep} to leave space around whole list

\newcommand{\lya}{Lyman-$\alpha$}

\newcommand{\kms}{km~s$^{-1}$}
\newcommand{\cmc}{cm$^{-3}$}
\apptocmd{\lim}{\limits}{}{}

\usepackage{hyperref}
\usepackage{xcolor}

\begin{document}
\title{A new 3D solar wind speed and density model based on interplanetary scintillation}
\author[0000-0001-8252-4104]{C. Porowski}
\affil{Space Research Centre PAS (CBK PAN),\\
Bartycka 18A, 00-716 Warsaw, Poland}
\author[0000-0003-3957-2359]{M. Bzowski}
\affil{Space Research Centre PAS (CBK PAN),\\
Bartycka 18A, 00-716 Warsaw, Poland}
\author[0000-0002-2982-1887]{M. Tokumaru}
\affil{Institute for Space-Earth Environmental research, Nagoya University, Nagoya, Japan}
\date{\today}

\begin{abstract}
The solar wind (SW) is an outflow of the solar coronal plasma, expanding supersonically throughout the heliosphere. SW particles interact by charge exchange with interstellar neutral atoms and on one hand, they modify the distribution of this gas in interplanetary space, and on the other hand they are seed population for heliospheric pickup ions and energetic neutral atoms (ENAs). The heliolatitudinal profiles of the SW speed and density evolve during the cycle of solar activity. A model of evolution of the SW speed and density is needed to interpret observations of ENAs, pickup ions, the heliospheric backscatter glow, etc. We derive the Warsaw Heliospheric Ionization Model 3DSW (WawHelIon 3DSW) based on interplanetary scintillation (IPS) tomography maps of the SW speed. We take the IPS tomography data from 1985 until 2020, compiled by \citet{tokumaru_etal:21a}. We derive a novel statistical method of filtering these data against outliers, we present a flexible analytic formula for the latitudinal profiles of the SW speed based on Legendre polynomials of varying order with additional restraining conditions at the poles, fit this formula to the yearly filtered data, and calculate the yearly SW density profiles using the latitudinally invariant SW energy flux, observed in the ecliptic plane. Despite application of refined IPS data set, a more sophisticated data filtering method, and a more flexible analytic model, the present results mostly agree with those obtained previously, which demonstrates the robustness of IPS studies of the SW structure. 
\end{abstract}

\section{Introduction}
\label{sec:intro}

The solar wind (SW), a supersonic stream of plasma expanding continuously from the Sun, creates a bubble-like region in interstellar matter filled with solar wind plasma, called the heliosphere. Solar wind ions and the neutral component of interstellar matter that penetrates inside the heliopause interact by charge exchange. This process operates both inside the solar wind termination shock and in the inner heliosheath, i.e., between the termination shock and the heliopause. The charge
exchange reaction products are energetic neutral atoms, which are former SW ions that were neutralized due to transfer of an electron from a hydrogen atom and maintain their velocity from the moment of charge exchange. Some of these former solar wind ions run towards the Sun, where they are detected by neutral atom detectors installed onboard Interstellar Boundary Explorer \citep[IBEX; ][]{mccomas_etal:09a}, thus bringing information about the physical state of the parent plasma in remote regions. However, to correctly interpret this information, one needs to know the input solar wind structure and its variation with time. Similarly, the knowledge of the solar wind structure is needed to account for ionization losses in interstellar neutral gas and energetic neutral atoms inside the heliosphere \citep[e.g.,][]{bzowski:08a,bzowski_etal:13a}. Knowledge of the solar wind structure is also crucial for many other areas of the solar system and space weather research.

SW properties can be determined from measurements {\emph{ in situ}}. Direct measurements of SW parameters have been performed in the ecliptic plane since the beginning of the space age and nowadays they are cross-calibrated and routinely published in the OMNI2 collection \citep{king_papitashvili:05}. However, only the Ulysses mission \citep{wenzel_etal:89a, bame_etal:92a, gloeckler_etal:92, mccomas_etal:08a} measured in situ the SW parameters in polar regions of the Sun. Ulysses operated in the years 1990--2009, so currently direct and continuous SW speed measurements at higher heliolatitudes are not available. While previously, the out-of-ecliptic structure of SW could be obtained in a few locations from images of cometary ion tails \citep{brandt_etal:75a}, currently the SW speed measurements at high heliolatitudes are performed using interplanetary scintillations (IPS) of radio waves observed on the Earth \citep[e.g.,][]{jackson_etal:97a,jackson_etal:98a,jackson_etal:11b,jackson_etal:15a,jackson_etal:20a,kojima:79a,kojima_etal:98a,kojima_etal:04b,kojima_etal:07a,manoharan:93b, manoharan:12a, mejia-ambriz_etal:15a,tokumaru_etal:00a,tokumaru_etal:10a,tokumaru_etal:12b,tokumaru:13a,tokumaru_etal:15a,tokumaru_etal:21a} and from spaceborne observations of the \lya{} helioglow \citep{lallement_etal:85a, summanen_etal:93, summanen_etal:97,bzowski_etal:03a, lallement_etal:10b, katushkina_etal:13a, katushkina_etal:19a, koutroumpa_etal:19a}.

\begin{figure}[!ht]
\center{
       \includegraphics[width=0.8\textwidth]{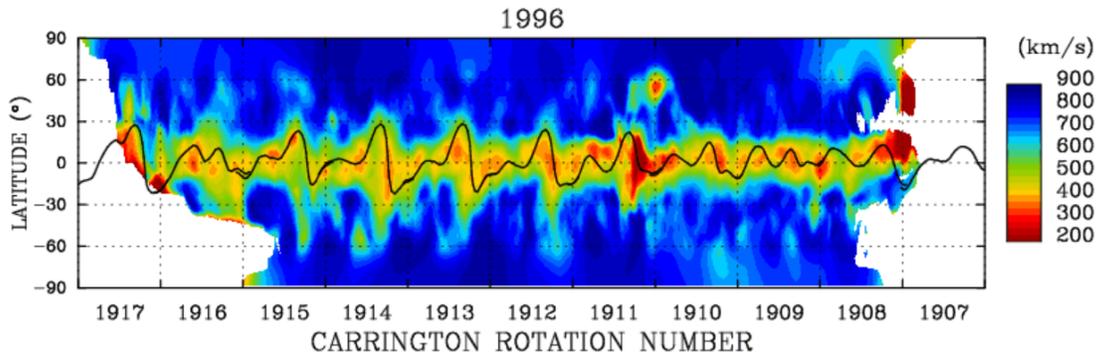}
       \caption{An example Carrington speed map for 1996. An extensive gap in the data at the southern hemisphere during the 1996 winter months is noticeable.
       }
\label{CarrSpeedMap}
}
\end{figure}

IPS is a random, sub-second fluctuation in the wave intensity of compact radio sources observed by radiotelescopes. They are caused by radio wave diffraction from small-scale ($\sim100$ km) electron-density irregularities in the SW. As the irregularities drift across the line of sight, rapidly changing diffraction patterns of radio waves are observed on the Earth as scintillations \citep{hewish_etal:64a}. IPS observations have been performed since the 1960's \citep{tokumaru:13a}. Assuming a frozen-in hypothesis, i.e., that the density irregularities in SW are drifting without change of their properties, and that the SW flows outward radially from the Sun, a cross-correlation analysis of multi-station IPS data allows for the reconstruction of the SW speed along the line of sight. Such speeds are affected by the so called line-of-sight integration effect, which blurs the fine structures. Using tomographic analysis \citep{kojima_etal:97a,jackson_etal:97a,jackson_etal:98a}, the IPS speeds along the line of sight can be deconvolved to a spatial speed distribution, with the line-of-sight integration effect removed, providing 3D Carrington synoptic speed maps of the SW. An example Carrington speed map is shown in Figure~\ref{CarrSpeedMap}.

These maps have inevitable gaps \citep[for a broader discussion of the gaps and methods of filling them, see, e.g.,][]{sokol_etal:15d}. However, for accurate studies of global heliospheric structure that can be extended to the outer heliospheric boundary, coverage that is homogeneous in time and heliolatitude is needed. To solve this problem in earlier analyses, the 3D Carrington speed maps are used to obtain average yearly latitudinal speed profiles on a fixed regular grid. This line of research is presented in \citet{sokol_etal:13a, sokol_etal:19a, sokol_etal:20a}. In these papers, the Carrington maps are longitudinally concatenated and averaged, and the resulting mean yearly speed profiles are approximated by an analytic function, which is subsequently mapped on a fixed heliolatitudinal grid for each year. The approximation function introduced by \citet{sokol_etal:13a} is a piecewise first- and second-order polynomial, where approximations of each yearly profile of the SW speed were adopted as piecewise functions composed of straight-lines in polar regions, flatly connected to smoothly-connected parabolas between the polar regions. The heliolatitude boundaries between the pieces (the straight lines and the parabolas) are adjusted to maximize the agreement with the data. To calculate the SW speed at a given heliolatitude for time moments between halves of individual years, a linear interpolation between the values obtained for this heliolatitude from the two adjacent yearly profiles is used. 

Recently, the entire set of IPS observations was reprocessed using a revised version of tomographic analysis \citep{tokumaru_etal:21a}. The reason for this revision was that discrepancy between IPS and in situ observations became prominent in Cycle 24: the SW speeds derived from the tomographic analysis of IPS observations were systematically higher (by about $71\pm26$ km/s) than those from the OMNI measurements for the period after 2009. This discrepancy was found to be improved by changing the value of the power law index $\alpha$ in the empirical relation $\Delta n_e\sim V^\alpha$ that is assumed in the tomographic analysis. Here, $\Delta n_e$ and $V^\alpha$ are the SW density fluctuations and speed, respectively. In the revised version of tomographic analysis, the index $\alpha$ was assumed to be variable, depending on the year. The variation in $\alpha$ is considered to be closely linked with the long-term decline in the solar activity. By allowing this change in the IPS analysis, the discrepancy between the IPS and in situ observations was reduced to $35\pm23$ \kms{} for the period after 2009, even though the tendency to overestimate the SW speed in the reprocessed data set still remains. 

Here, we overhaul the approach originally defined by \citet{sokol_etal:13a}. We take the most recent version of the IPS SW speed maps \citep{tokumaru_etal:21a}, we introduce a novel method of filtering these maps against outliers, and we use a different approximating function, based on Legendre polynomials. The model of yearly-averaged SW speed and density is extended until 2021. 

The methodology of our SW speed modeling consists of two main parts: initial processing of the IPS-derived Carrington speed maps and searching for the best yearly fits to longitudinally averaged yearly speed maps. The main goal of the initial processing is to provide filtered, averaged and binned yearly latitudinal speed profiles of SW without bias. Bias removing is done by filtering out the map background before averaging the speed maps. The filtering method is based on statistical analysis to avoid any arbitrary cuts.

After filtering, the Carrington speed maps are averaged in the selected bins over CRs to provide the Carrington-mean latitudinal SW speed profiles. The binning is done for each of the CRs with IPS data. The binned data are then used for fitting an approximating function to the yearly SW speed data. The approximation model we use is more flexible than that used by \citet{sokol_etal:20a}. The speed profiles are defined as sums of Legendre polynomials with appropriate coefficients, with an additional requirement that the derivative of the profile over the pole is 0. The range of the model used varies between the years, depending on the data. The fitting is done simultaneously to all Carrington-averaged profiles from a given year. Each year is fitted separately. Here again, to eliminate arbitrary decisions, during the fitting procedure we apply a rigorous statistical analysis to identify the optimum range of the model for the given yearly SW speed profile model. 

In the final step, an analysis of differences between the model and OMNI in situ measurements is performed and the yearly model speed profiles adjusted, similarly as it was done by \citet{sokol_etal:20a}. The SW density for a given year is calculated based on the adjusted yearly speed profile and the magnitude of the latitudinally invariant SW energy flux for the given year \citep{leChat_etal:12a}.

The input data are presented in Section~\ref{sec:dataGeneral}. The data filtering method is presented in Section~\ref{sec:dataCleaning}. There, we estimate the mean bias as a function of latitude and describe its properties. Subsequently, we present the results of filtering the data in Section \ref{sec:initialProcessing}. The approximation model is presented in Section \ref{sec:approxModel}. The final SW speed binning and model fitting is discussed in Section \ref{sec:speedFitting}. Then, the SW speed model is adjusted to the OMNI2 speeds in the ecliptic plane (Section \ref{sec:OMNICorr}). The results are discussed in Section~\ref{sec:SWspeedresults}, and the density model is calculated using the SW energy invariant measured {\emph{in situ}}, as presented in Section~\ref{sec:density}. The results are discussed in Section~\ref{sec:discussion}, and we finish with summary and conclusions (Section \ref{sec:conclusions}).

\section{Data}
\label{sec:data}
\subsection{General description of the data}
\label{sec:dataGeneral}

As input to the analysis, we use the most recent IPS Carrington speed maps publicly available from \href{https://stsw1.isee.nagoya-u.ac.jp/ips_data-e.html}{the Institute for Space-Earth Environmental Research (ISEE) at Nagoya University, Japan} \citep{tokumaru_etal:10a,tokumaru_etal:12b, tokumaru_etal:21a}. These maps were derived using the Computer Assisted Tomography (CAT) method applied to the ISEE IPS observations \citep{tokumaru:13a}, covering the entire interval of available observations from 1985 through 2020, with a gap in 2010 because of antenna system modification. These maps are available on a regular 1\degr{} by 1\degr{} mesh in heliographic coordinates. The maps are provided in time intervals corresponding to individual Carrington rotations of the Sun. 

Because of snow loading on the two movable antennas in winter, the IPS velocity data collection is stopped from December to February. In effect, regular gaps in the IPS observations are present in each yearly Carrington speed maps for the CRs during this period, and the available number of CRs in the speed map for a given year is about 11; the boundaries of Carrington periods do not coincide with the beginning of the year, and thus the gap in data coverage over the winter months is variable.

In addition to these regular gaps, some additional gaps may appear due to observing conditions and the sparsity of the available radio sources in the sky. The ISEE IPS data are collected using a UHF radio-telescope array, which consists of antennas operating at a frequency 327 MHz. The data are collected from observations of 15--20 compact radio-sources (mostly quasars, observed daily at small angular elongations from the Sun). In winter months, the Sun is low on the horizon, and very few sources are able to be viewed south of the Sun at this time. Thus, sources in the regions southward of the Sun are not always sufficient to provide a uniform coverage of the Carrington speed map in the southern hemisphere. This factor also affects the mean speed profiles, because a lower number of data points may cause a larger scatter of binned speed points near the south pole of the Sun.

The irregular distribution of radio-sources in the sky is one of the reasons why the accuracy of SW speed reproduction varies within the Carrington maps. In particular, when very few radio sources are available in a large region in the sky, the tomography analysis sometimes returns outlying very low or very high speeds. In the original analysis by \citet{tokumaru_etal:21a}, the boundaries of allowable speed values were set to 200~\kms{} at the lower end and 850~\kms{} at the high end of the speed range.

Based on comparison of IPS-derived and Ulysses in situ observations, \citet{sokol_etal:13a} noticed that generally, individual monthly Carrington maps with less than 30\,000 data points have a lower accuracy than the maps with a larger number of points, and to prevent a bias, they rejected these sparsely-populated maps from further analysis. But these sparsely-populated maps may include well-populated regions that can safely be used in the analysis. Therefore, we devised a data filtering method based on statistical search for outliers and employed it as the first step in our data processing. This method and its results are presented in the next section.
 
\subsection{The need for data filtering}
\label{sec:dataCleaning}
\begin{figure}[!ht]
\center{
       \includegraphics[width=0.5\columnwidth]{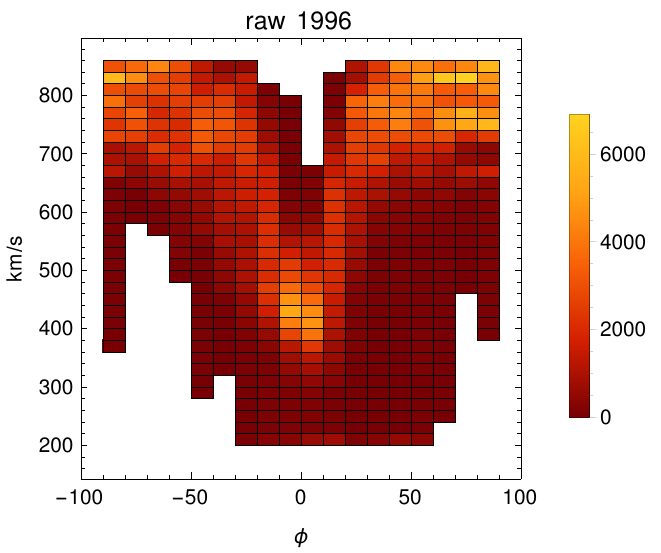}
       \caption{Example density histogram of speed points taken from the 1996 Carrington speed map of Figure \ref{CarrSpeedMap}, plotted as a function of heliolatitude. The map background, which covers speeds from 200--850 \kms{} in an extensive range of heliolatitudes, is clearly visible.
       %obrazek: Model2020\_Wizualizacja\_i\_Analiza.nb
       }
\label{hist1}
}
\end{figure}

To avoid bias arising due to the map background, we treat the Carrington maps as raw data, and we introduce additional processing of the maps before averaging. The need for data filtering is clearly seen in a density histogram of SW speeds, taken from a yearly Carrington map concatenated over heliolongitude, presented as an example in Figure \ref{hist1}. In the density histogram of speeds the majority of speed points generally follow a latitudinal V-shaped profile, which is a characteristic solar wind feature during solar minima, like in the example year 1996. But some of the speed points fall far outside the main V-shape area. Some of them reach the low-speed limit at 200 \kms, even though the majority of speed values at high heliolatitudes are about 800 \kms. The low speed values at higher heliolatitudes are not likely to be realistic during solar minimum, and will be referred to as the map background. It must be stressed here that some of the spread in the yearly-accumulated data is due to physical variation in SW speeds. 

To identify and reject outlying data points, we used a novel statistical method based on the Generalized Extreme Studentized Deviate (ESD) Many-Outliers detection procedure by \citet{rosner:83a}.
In this method, it is assumed that the data sample is a mixture of a certain number of outlier points in addition to the ``correct data'' population conforming with the normal statistical distribution.
This  method hypothesizes that the sample includes up to $n$ outliers and assumes that they deviate greatly from others in the sample. We start by selecting $n$ points the most distant from the median of the sample. We calculate the test statistics and critical values for each of the suspect-outlier points from 1 to $n$ and reject the point for which the test statistic value is higher than the critical value, obtained from the Student T distribution. We reiterate this procedure: after each step, the $i$-th observation is rejected, and the statistics and the critical value are recalculated. The procedure returns the number of outliers, which is the largest $i\in n$ at which the $i$-th test statistics is larger than $i$-th critical value.

\begin{figure}[!ht]
\center{
       \includegraphics[width=0.5\columnwidth]{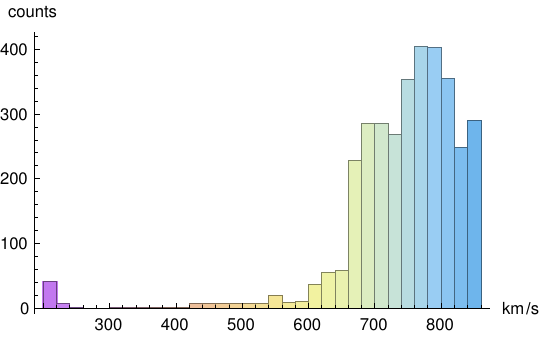}
       \caption{An example histogram of speed values of points belonging to the equatorial $1\degr${} bin at the heliolatitude $36\degr$N. The data points are taken from the 1996 Carrington map.
       %obrazek: Wizualizacja i Analiza
       }
\label{hist3}
}
\end{figure}

\subsection{Filtering Carrington speed maps against background}
\label{sec:initialProcessing}
The analysis begins with application of initial map filtering procedure, which identifies and removes speed points belonging to the map background. The procedure is performed separately within individual heliolatitude bins, separately for each year, but the speed points from the whole year are analyzed simultaneously. Also, each heliolatitude bin is analyzed independently.

The filtering procedure is based on the fact that the map background points feature a large deviation from the median value of the distribution in the respective heliolatitude bin. We use the median value because the speed distributions are strongly asymmetric, as shown in the Figure~\ref{hist3} example. Not all speed distributions in heliolatitude bins are asymmetric but the use of median is not a problem in these cases. We assume that the background points are outliers of the speed distribution around the median value, so the map filtering relies on identification and removing of the outliers in each bin.

During the filtering, we operate on $1\degr${} heliolatitude bins, coincident with the maximum resolution of the original Carrington maps. In each $1\degr${} heliolatitude bin, we use all speed values in a given calendar year, without dividing them into individual CRs. Although we use narrow bins, the number of points in each $1\degr${} heliolatitude bin is large. This is because the speed points in the Carrington maps do not represent the measured speed values, but they are a result of CAT reconstruction on a dense grid of heliolongitude and heliolatitude. Therefore, we treat the points in each heliolatitude bin as oversampled data, which in general follow the Student T-like distribution.

To identify and remove the outliers, we transform the speed distributions by calculating absolute values of speed differences between individual speed points and the median speed within a given bin. Next, we search for the maximum number of outliers in the distribution of absolute speed differences, using the one sided extreme Studentized distribution (ESD) test \citep{rosner:83a}. After identification of the outliers, they are removed from the Carrington maps before averaging.

Even though for a given year the speed points from all CRs are analyzed at once, the information about CR location for each point is preserved and used subsequently during averaging. Since we operate on the original-resolution Carrington speed maps, we need to process 180 $1\degr${} equatorial bins per year. The central positions of the latitudinal $1\degree${} bins correspond to the original resolution of the Carrington speed map. This processing is time consuming because the maps consist of a large amount of speed points.

No initial averaging of any parts of the maps is performed. The sharp edges of the blank areas on the maps precludes this because averaging in the presence of such edges would result in reduction of speed values near the blank areas. Proceeding with the maximum resolution speed maps is done to preserve as many points as possible, since the data have extensive gaps. In this way, we can use speed points at the south pole belonging to CRs with extensive gaps of data. Preserving this information is important because of generally poor map coverage in the southern hemisphere is motivated by our intent to sustain the available statistics as much as possible. For example, for CR1908 in the yearly map presented in Figure \ref{CarrSpeedMap}, the southern area is poorly covered, but the speed points existing in this CR are used, while in previous models such CRs would be rejected entirely. 

\begin{figure}[!ht]
\centering
\includegraphics[width=0.65\columnwidth]{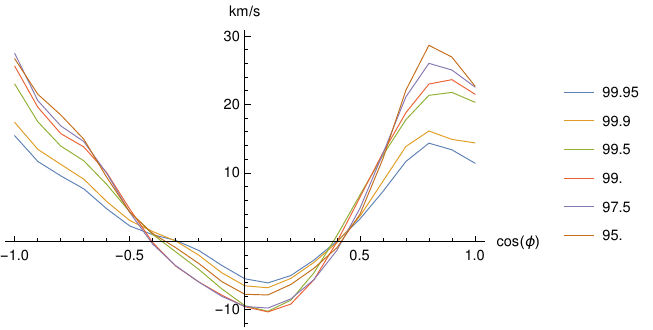}
    \caption{Mean values of the correction obtained as a result of filtering of the Carrington speed maps for the entire data set, plotted as a function of $\cos(\pi/2-\phi)$ for typical values of CL, listed in the figure legends.
	}
\label{Corr}
\end{figure}

The filtering method is governed by only one parameter, i.e., the confidence level (CL). We use the same CL for all histograms in the  analyzed data set, but the optimum CL value to use was subjected to further analysis. Different values of CL will effect in different intensity of map filtering. The optimal filtering intensity, driven by the CL value, is determined as follows. For each year, we calculate the mean SW speed profiles for the map with background (i.e., for not filtered). Subsequently, we repeat the calculations for the map with the background removed, using several typical values of the CL as a parameter of the one-tailed statistical test: 95\%, 97.5\%, 99.0\%, 99.5\%, 99.9\%, 99.95\%. Then we compare how the correction profile averaged over all years changes after filtering depending on the CL used. The magnitudes of such changes as a function of latitude averaged over all years can be regarded as a typical correction of the latitudinal profiles, obtained with respect to the profiles modeled without map filtering (see Figure~\ref{Corr}). The behavior of the this correction for different levels of the filtering intensity allows us to select the optimum CL.

This is done by inspection of the mean correction profiles for different CLs. We found that the correction profiles have smooth shapes, and are similar for different CLs. An evolution of the correction profile with CL value is shown in Figure~\ref{Corr}. Clearly, the modifications of a yearly profile due to filtering for various CLs has similar features. The interpretation of the typical correction is that on average, the map filtering increases the speed values by $\sim25$ \kms{} at both poles. This means that typically, the bias due to the map background causes an underestimation of speeds at the poles. At the equator, the filtering results in decreasing  the modeled speed by $\sim10$ \kms. This is important also in the context of differences between the IPS and in-situ results in the ecliptic plane, discussed in Section~\ref{sec:OMNICorr}. The typical correction is north-south asymmetric. The asymmetry may arise because of non-uniform statistics and data coverage at the polar regions. 

The filtering results in a systematic modification of speeds at all heliolatitudes. At the equator, probably the adopted lower speed limit of 200 \kms{} slightly biases the mean profile gathered from the raw maps. The latitudes at which the speed limits do not introduce any difference in the mean sense are around $\pm25\degr$, but this may vary between the individual years. It is also seen that for the CL 99.5\% and less, a stabilization of the mean correction value occurs. This behavior suggests that, starting from a certain value of CL, the filtering method becomes stable for smaller CL values, otherwise the mean bias correction profile would increase with the CL decrease. This also confirms that the mean profile distortion, caused by the map background and the 200 \kms{} speed limit, is compensated for, since further elimination of outliers with a smaller CL brings negligible effect. However, the smaller CL, the lower deviation of points around the model fit, which leads to an underestimation of the accuracy of the model. Therefore, to avoid this effect, in further analysis we use CL $=99\%$, i.e., the largest CL at which the mean bias correction profile becomes stable. 

\begin{figure}[!ht]
\center{
       \includegraphics[width=0.5\columnwidth]{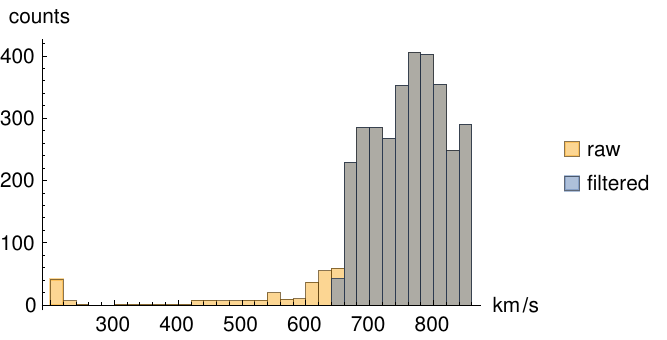}
       \caption{Example histograms of speed values of points belonging to the $1\degr${} bin at the heliolatitude $36\degr$N. The data points are taken from the 1996 Carrington map before and after filtering. The raw histogram is identical to that shown in Figure~\ref{hist3}.
       }
\label{hist4}
}
\end{figure}

\begin{figure}[!ht]
\center{

    \includegraphics[width=0.45\columnwidth]{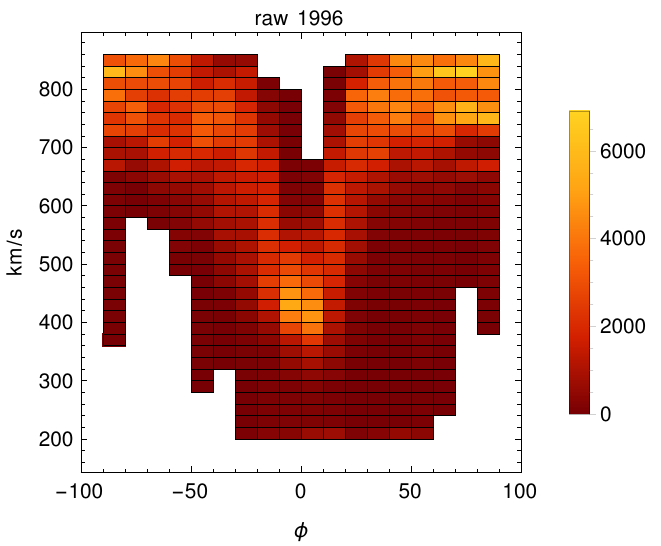}
\includegraphics[width=0.45\columnwidth]{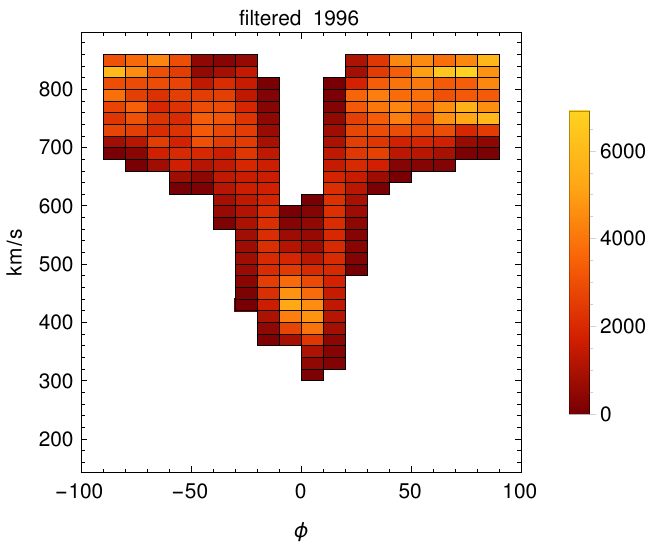}
       \caption{Example density histograms of speed points taken from the 1996 Carrington speed map before (left) and after (right) filtering, plotted as a function of heliolatitude. Extensive elimination of the map background due to filtration is visible.
       %obrazek: Model2020_Wizualizacja_i_Analiza.nb
       }
\label{hist2}
}
\end{figure}

With this, after identification and rejection of the map background, we obtain filtered Carrington speed maps, which are used for further steps in the model construction. The average yearly latitudinal profiles from these maps have a reduced or eliminated bias. The effect of filtration can be appreciated in Figure~\ref{hist4}, where the histogram of the data from the raw map from 1996 is compared with the histogram of filtered speeds, and in Figure \ref{hist2}, where heliolatitude histograms of speeds before and after filtering are presented. The value of correction for individual years increases with heliolatitude, with extreme values from $-35$~\kms{} in the equatorial area to $78$~\kms{} at higher latitudes.

\subsection{Preparation yearly profiles for fitting}
\label{sec:yearly4fitting}

Next, to prepare the yearly data for speed profile fitting, we change the domain of the Carrington speed maps from heliolatitude ($\phi$) to $\cos(\pi/2-\phi)$, because we want to model the speed profiles using orthogonal functions at the $\langle-1,1\rangle$ interval. A division of the domain into bins with the same width yields equi-areal latitudinal bins, which means that the bin widths in degrees towards the solar poles are wider than those at the equator. The widening of the bins is a desirable feature, because the number of available points towards the poles decreases due to the lack of IPS data. With the equi-areal bins, the statistics in the bins is more homogeneous, and will result in reduction of fluctuations at the poles in the fitted models. An empirical study showed that the optimal latitudinal bin width is 0.05, i.e., we use 40 bins in the entire latitudinal $\langle-1,1\rangle$ range. We found that a change of the bin width by a factor of 1.5 does not substantially affect the results.

\begin{figure}[!ht]
\center{

     \includegraphics[width=0.5\columnwidth]{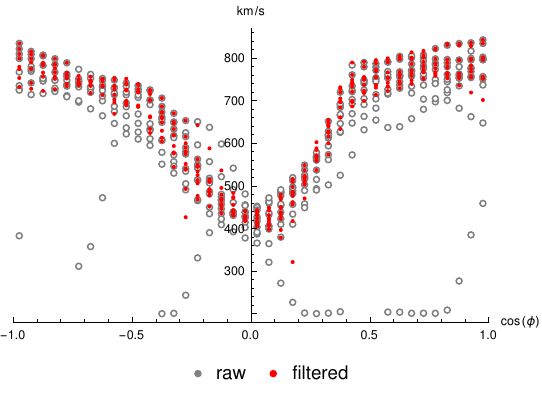}
      \caption{
	       An example plot of the binned speed values for the raw and the filtered Carrington speed map from 1996. The entire range of $\cos(\pi/2-\phi)$ is presented, as it is used for modeling of the mean yearly profile. The red points are used for fitting, the gray points have been filtered out.
       %obrazek: Model2020_Wizualizacja_i_Analiza.nb
       }
    \label{binnedAll}
    }
\end{figure}

\begin{figure}[!ht]
\center{

    \includegraphics[width=0.5\columnwidth]{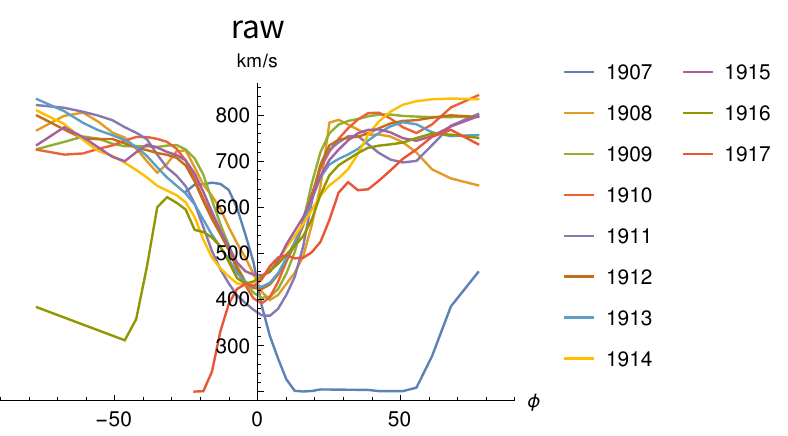}
    \includegraphics[width=0.5\columnwidth]{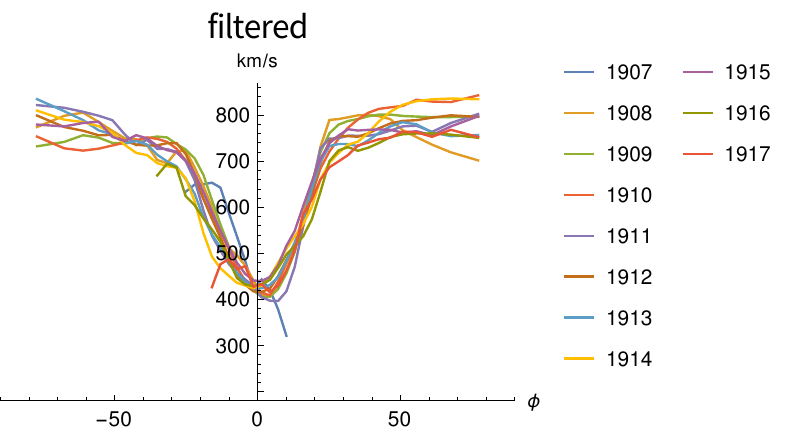}
      \caption{
	     An example plot of binned speed values for the raw and the filtered Carrington speed maps for 1996. Individual CRs with binned data are color-coded.
	       }
\label{binnedCR}
}
\end{figure}

Subsequently, we average the speed points within the bins, separately for each CR. The speeds in each bin are averaged selectively, after being gathered from the CR to which they belong. So, for a given year, we obtain as many binned profiles, as there are CRs available that year. This is different compared with the processing performed during the filtering. During the filtering, all speed points from the entire year were processed together, but now data from individual CR are treated separately by using identical bins. This binning provides profiles for all CRs for which the data are available in the original data set. The data prepared in this way from all CRs for a given calendar year are used in the fitting of yearly model profiles. Some of the profiles, however, have gaps in the latitudinal coverage, and can be dealt with as demonstrated below.

An example visualization of binned points, ready to fit, is shown in Figure \ref{binnedAll}. A comparison of binned profiles for the raw and the filtered Carrington maps are presented in Figure \ref{binnedCR}. The data in Figure~\ref{binnedCR} are the same as in Figure~\ref{binnedAll}; the binned profiles are shown with their CR numbers, to visualize the impact of map filtering. It is seen that after filtering the mean CR profiles are less noisy, but nevertheless a certain noise still persists. Clearly, the filtering removes many of the profile distortions and artifacts.

An example of the distortion removal can be appreciated in the profile from CR1917. This profile was obtained at the end of the IPS data collection period in 1996. The lack of sufficiently abundant IPS data in this CR caused that the profile is shifted down in the northern hemisphere, and is very distorted in the southern hemisphere. The filtering procedure removes the distortion, and moves the profile into an adequate place from the statistical point of view for the assumed CL. That in the case of extreme distortions the profiles are just cut out means that no extraction of information about the actual shape of the profile was possible because of insufficient data amounts. 

To prepare data for fitting yearly model speed profiles, we do not average the profiles from individual CRs over a calendar year. This is to preserve the SW speed variations between the CRs during a given year. We do so as a result of an empirical study that we performed, which showed that the point distribution between individual CRs make the fitting procedure more efficient and reliable. The yearly models are fit to binned heliolatitude profiles for individual Carrington rotations for a given calendar year.

The final step of profile preparation for fitting is averaging the speed within the 40 bins.

\section{Modelling of the solar wind speed}
\label{sec:modeling}

\subsection{Approximation model}
\label{sec:approxModel}

The modeling is done using averaged speed profiles like that showed as an example in Figure \ref{binnedAll}. We model the SW speed profiles as a sum Legendre polynomials. These polynomials have the necessary flexibility needed to reproduce the various shapes of the profiles we need to approximate. However, tests showed that an additional condition of nulling the derivatives of the model at the poles is necessary to provide a general stability of the model and to avoid the propagation of the effects of low coverage of the polar regions into lower latitudes. Without these conditions, an unrealistic behavior of the fits at the poles sometimes occurs. Also, the use of binned speed values contributes to instability of the unconditioned model at the poles because of the lack of binned values exactly at the poles, i.e., at $\cos(\pi/2-\phi)=\pm1$. For this reason, we set the first derivatives of the Legendre polynomials at the poles to zero.

Denoting $z=\cos(\pi/2-\phi)$, a general form of the model is the following:
%\systeme{x_1=2r + s -t,x_2= r, x_3=-2s +2t, x_4=s, x_5=t}
\begin{equation}
V(z)=\sum_{i=0}^NQ_iP_i(z),
\label{eq:modelLegendre}
\end{equation}
with the conditions at the poles (i.e., at the points where $z=\pm1$):
\begin{equation}
\frac{dV}{dz}\bigg |_{z=\pm 1}=0,
\label{eq:modelLegendreDeriv}
\end{equation}
where $P_i$ is Legendre polynomial of $i$-th order, $Q_i$ is the coefficient of the $i$-th polynomial, and $N$ is the approximation order. 

Applying these conditions eliminates two of the free parameters $Q_i$. The formulae for $V(z)$ used in the fitting are algebraically rearranged by deriving a linear combination of the Legendre polynomials for a given $N$, in which the two coefficients standing at the highest orders of the Legendre polynomials are eliminated. The remaining free coefficients are obtained from the fit to the data. Derivation of the formulae was done using algebraic manipulation and differentiation in Wolfram Research
Mathematica. Legendre polynomials were obtained from Mathematica function LegendreP. The analytic formulae are too large to show them here, especially since they change from one order to another with the change of the order used because of the additional condition specified in Equation~ \ref{eq:modelLegendreDeriv}.

\begin{figure}[!ht]
\centering
\includegraphics[width=0.5\columnwidth]{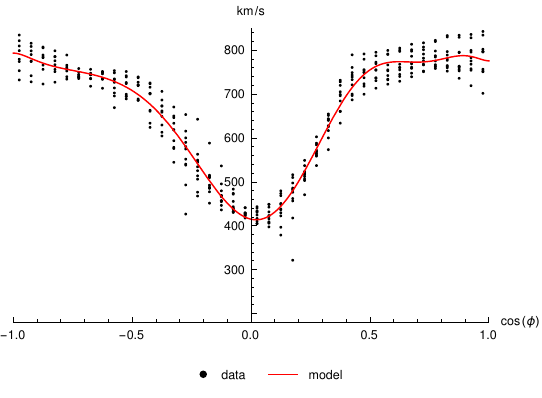}
    \caption{The final fit for 1996, before applying the OMNI2 adjustment. The black points are the binned speed values for the filtered Carrington map for 1996. The red line is the speed profile modeled using Legendre polynomials (Equations \ref{eq:modelLegendre} and \ref{eq:modelLegendreDeriv}).
    }
\label{fig:FIT}
\end{figure}

\subsection{Fitting yearly model speed profiles}
\label{sec:speedFitting}
The least-squares fitting was performed using Mathematica function LinearModelFit. Wolfram Research  Mathematica version 12 was used. 

The mean SW speed changes from year to year. Depending on the year and the solar cycle phase, the complexity of the SW speed profile shapes varies. Since the Legendre polynomials are very agile, using too high an order in the fitting procedure may result in reproducing the remaining small data fluctuations in the model.  The goal here is to  avoid overfitting, i.e., in reproducing some remnant random variations in the data. In the case of the former model by \citet{sokol_etal:20a}, this issue was absent because of the simplicity of the approximating function (low-order polynomials). However, the same factor may have prevented that model from reproducing possible legitimate variations of solar wind speed with heliolatitude, like those proposed by \citet{katushkina_etal:13a, katushkina_etal:19a}, who reported that the solar wind flux maximizes at mid-latitudes in the northern and southern hemispheres based on analysis of the helioglow distribution in the sky.

Our analysis showed that models of different orders are optimal in different years. The choice of the optimal polynomial order is done for each year separately, based on statistical analysis of the fit residuals. We set the lower order of the Legendre polynomials to 8, and the upper one to 20, and we assumed that the optimal order of the Legendre polynomial should return residuals with the distribution characterized by the highest p-value of the Pearson $\chi^2$ test of normality.

However, using the ordinary $\chi^2$ measure was abandoned because of the lack of the uncertainties for individual speed points in the input maps. Using the dispersion in individual bins as an estimator of the uncertainty measure, which might be used in the ordinary $\chi^2$ measure, led to wrong results. This can be understood when one realizes that the input data points in high-resolution Carrington maps are results of mapping the results of CAT analysis on the coordinate grid. Therefore, the data points are not independent and may feature correlations within a certain spatial range in the maps. An example of such correlation can be seen in Figure~\ref{binnedAll}, where among the gray points (filtered out from the sample), series of data points forming continuous gradients are visible. This illustrates that the CAT method used to derive the map can introduce correlations between nearby data points. 

An example best fit for 1996 is shown in Figure \ref{fig:FIT}. Having fit the SW speed model for all years, we performed a reverse transformation of latitude from $\cos(\pi/2-\phi)$ to $\phi$, and we apply the OMNI2 adjustment (see Section \ref{sec:OMNICorr}).

\subsection{The OMNI2 adjustment}
\label{sec:OMNICorr}
A good agreement between IPS CAT speeds and Ulysses measurements was reported by \citet{tokumaru:13a}. However, in certain time intervals, systematic differences between IPS-derived speeds and those measured {\emph{in situ}} and compiled within the OMNI2 collection \citep{king_papitashvili:05} have been noticed \citep{sokol_etal:20a, tokumaru_etal:21a}.

\begin{figure}[!ht]
	\centering
	\includegraphics[width=0.5\columnwidth]{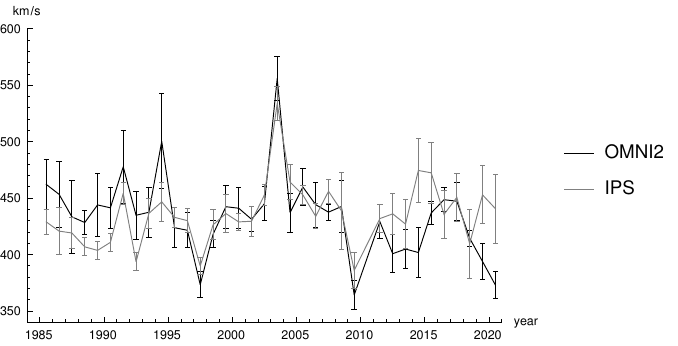}
	\caption{A comparison of the mean yearly speeds obtained from filtered Carrington maps at the solar equator with the yearly averaged OMNI2 speeds. The magnitudes of the uncertainty bars are equal to $1\sigma$ taken from the standard deviation of the CR-averaged speeds at the equator.}
\label{IpsOmniComp}
\end{figure}

Differences between the SW speeds obtained from the ISEE IPS measurements and Ulysses data were also studied in detail by \citet{sokol_etal:13a}, where the overall usefulness of the IPS measurements at all latitudes was demonstrated for the years of the Ulysses mission. However, \citet{sokol_etal:20a} reported that the mean yearly IPS speeds at the equator taken directly from the Carrington maps showed a divergence $\Delta v_{IPS-OMNI} = v_{IPS} - v_{OMNI}$ after 2011, i.e., after the year of the ISEE IPS antenna system upgrade. For the revised up-to-date IPS speeds, a comparison of the IPS speeds with OMNI2 time series is presented in Figure \ref{IpsOmniComp}, where the difference, previously reported by \citet{sokol_etal:20a} for the previous version of IPS speed maps, is generally reduced, but is still present. \citet{tokumaru_etal:21a} present a detailed study of the differences between the SW speed derived from IPS tomography and in situ measurements performed by Parker Solar Probe and those available in the OMNI2 collection on a much finer time scale than we have in this paper.

\begin{figure}[!ht]
\centering
\includegraphics[width=0.75\columnwidth]{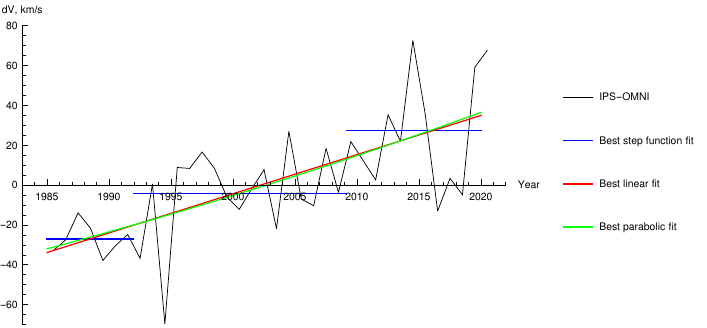}
    \caption{Time evolution of $\Delta v_{IPS-OMNI}$ in \kms{} at the solar equator.
    Best fits of the linear, parabolic, and step functions are plotted.}
    %/home/cporowski/IPSProfilesUpdate/Model2020A/Final/StepFunction.nb
    \label{IPStimev}
\end{figure}

We studied in detail the $\Delta v_{IPS-OMNI}$ time series. Although a look at the time evolution of this quantity may suggest it features a simple linear trend (see Figure~\ref{IPStimev}), we performed a statistical analysis of the goodness of fit for several different functions. In addition to a linear trend, parabolic and step functions for the fit to the time series of $\Delta v_{IPS-OMNI}(t)$ were studied. The boundaries in the step function were used to quantify a hypothetical coincidence between the antenna setup and changed the level of agreement of the IPS data with the OMNI2 measurements.
The better agreement between the IPS and the OMNI2 data that exists during the years of the Ulysses mission, when \it in situ \rm measurements were used for the IPS calibration of raw data, is another reason to consider the use of a step function to derive the adjustment for the IPS yearly profiles.

The three alternative models have different numbers of free parameters. We performed the fits to the three functions using Wolfram Mathematica and checked if the use of a model with larger parameter numbers yielded a statistically significant improvement of the model. We found that the values of the Akaike Information Criterion (AIC) for the three models are similar, and none of the models seems to be significantly favored from the current statistics.

A possible explanation for the behavior of the systematic differences in time was pointed out by \citet{tokumaru_etal:21a} who suggested that it might be due to the effect of time variation of the index $\alpha$, as mentioned in Section \ref{sec:intro}. The physical property of SW microturbulence is likely to evolve differently from the past cycles in accompany with the significant weakening of the solar activity for Cycle 24. According to the earlier study \citep{tokumaru_etal:18a}, a pronounced growth of the low-$\Delta n_e$ stream occurred for the slow SW in Cycle 24. This may result in a drastic modification of the $\Delta n_e$-$V$ relation used in IPS SW speed reconstructions, particularly for the slow SW. The IPS-OMNI comparison is considered to be seriously affected by this change, because the OMNI data were collected in the ecliptic plane, where the slow SW dominates. The change in the $\Delta n_e$-$V$ relation revealed here may suggest that the fractional density fluctuations $\Delta n_e/n_e$ became more dependent on $V$ in this cycle than before, providing an important insight into the solar wind acceleration mechanism. While this approach resulted in a reduction of $\Delta v_{IPS-OMNI}$ in comparison with these reported by \citet{sokol_etal:20a}, the present version of the IPS Carrington speed maps does not eliminate them altogether. 

A detailed look at individual years on the $\Delta v_{IPS-OMNI}$ plot shows a large difference appeared for year 1994, which was not present before. Additionally, a small systematic difference is visible for the years 1985--1992. In this period, all values of the IPS speeds are slightly lower than the OMNI2 speeds, which was not present in the previous version of the IPS data. However, taking into account the uncertainty, the differences seem to be reduced when compared with the previous version of the IPS speeds, but the divergence after 2011 is still noticeable. Almost no divergence is observed during the years of the Ulysses mission.

Since the OMNI2 time series is compiled from many different experiments and measurement methods, which are carefully curated \citep[see also on-line description of the data scaling and calibration at  https://omniweb.gsfc.nasa.gov/html/ow\_data.html]{king_papitashvili:05}, we decided to adjust the model derived from the IPS Carrington maps so that it agrees with the OMNI2 collection {\emph{ in situ}} time series. 

After a detailed study of the $\Delta v_{IPS-OMNI}$, taking into account that the considered differences coincide with the above-mentioned variations of $\Delta v_{IPS-OMNI}$ in different periods of data set, and having considered several potential methods of the adjustment derivations, we concluded that the optimal strategy for adjusting the IPS data was the method proposed by \citet{sokol_etal:20a}. Thus, we calculate the difference $\Delta v_{IPS-OMNI}$ separately for each year and subtract it from all latitudinal bins. We found that only 9\% of the adjustments are statistically significant, i.e., larger than the joint uncertainties of the yearly averages of the OMNI2 and IPS speeds. 

For the entire data set, this adjustment provides very small differences for the years that show a good agreement with the OMNI2 time series, while it reduces large values of $\Delta v_{IPS-OMNI}$ for the years when these differences are large, like 1994, 2014, and 2020. The magnitudes of the adjustment values for individual years are listed in Table~\ref{tab:omniCorrections}. 
\begin{table}
    \caption{The list of applied OMNI2 adjustment values.}
    \label{tab:omniCorrections}
\centering
    \small
    \begin{tabular}{l c | l c | l c | l c | l c }
        year & $-\Delta v_{IPS-OMNI}$ & year & $-\Delta v_{IPS-OMNI}$ & year & $-\Delta v_{IPS-OMNI}$ & year & $-\Delta v_{IPS-OMNI}$ & year & $-\Delta v_{IPS-OMNI}$ \\
        \hline
        1985.5 &  32.8545  & 1986.5 &  27.4526  & 1987.5 &  13.8923  & 1988.5 &  21.6139   & 1989.5 &  37.7993\\
        1990.5 &  30.491   & 1991.5 &  24.7447  & 1992.5 &  36.7307  & 1993.5 &  -0.589545 & 1994.5 &  69.6817\\
        1995.5 &  -8.87191 & 1996.5 &  -8.39963 & 1997.5 & -16.6271  & 1998.5 &  -8.41365  & 1999.5 &   6.00502\\
        2000.5 &  12.1142  & 2001.5 &   1.81115 & 2002.5 &  -7.78424 & 2003.5 &  21.8825   & 2004.5 & -27.0065\\
        2005.5 &   6.95075 & 2006.5 &  10.3166  & 2007.5 & -18.4858  & 2008.5 &   3.65364  & 2009.5 & -21.8524\\
        2011.5 &  -2.67103 & 2012.5 & -35.2988  & 2013.5 & -22.0815  & 2014.5 & -72.6916   & 2015.5 & -35.5581\\
        2016.5 &  12.9614  & 2017.5 &  -3.42591 & 2018.5 &   4.8346  & 2019.5 & -59.2082   & 2020.5 & -67.6243\\

    %    1985.5& 1986.5& 1987.5& 1988.5& 1989.5& 1990.5& 1991.5& 1992.5\\
    %    1993.5& 1994.5& 1995.5& 1996.5& 1997.5& 1998.5& 1999.5& 2000.5\\
    %    2001.5& 2002.5& 2003.5& 2004.5& 2005.5& 2006.5& 2007.5& 2008.5\\
    %    2009.5& 2011.5& 2012.5& 2013.5& 2014.5& 2015.5& 2016.5& 2017.5\\
    %    2018.5& 2019.5& 2020.5&  &&&&\\
    \end{tabular}
    \small
\end{table}

\newpage

\begin{figure}[ht]
\centering
 \includegraphics[width=0.33\columnwidth]{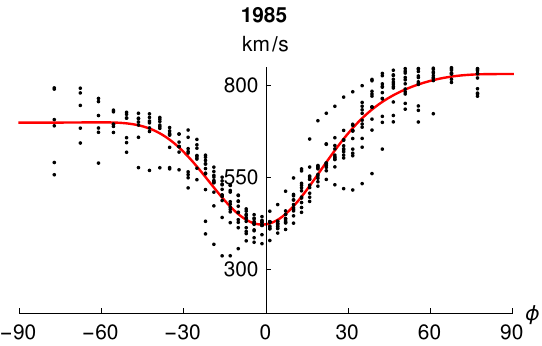}
 \includegraphics[width=0.33\columnwidth]{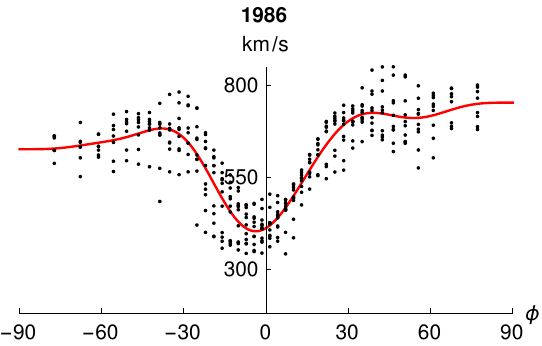}
 \includegraphics[width=0.33\columnwidth]{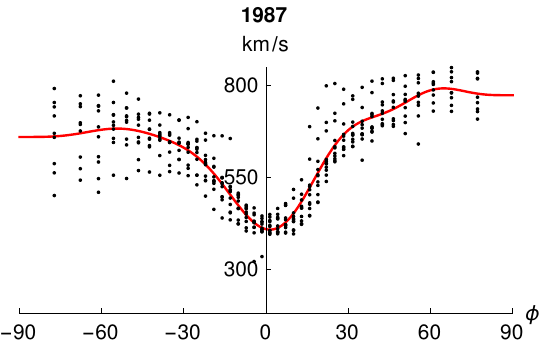}
 
 \includegraphics[width=0.33\columnwidth]{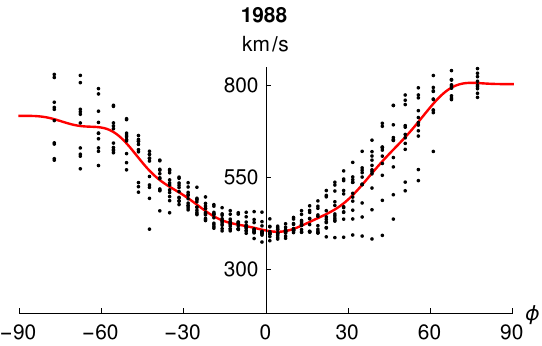}
 \includegraphics[width=0.33\columnwidth]{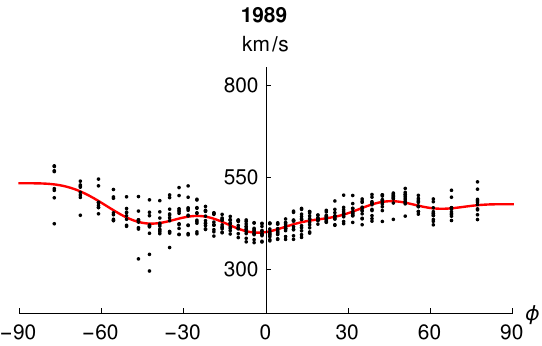}
 \includegraphics[width=0.33\columnwidth]{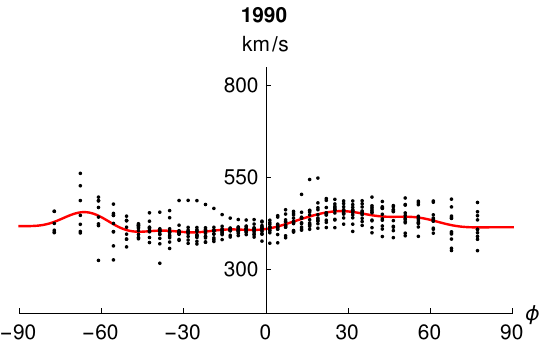}
 
 \includegraphics[width=0.33\columnwidth]{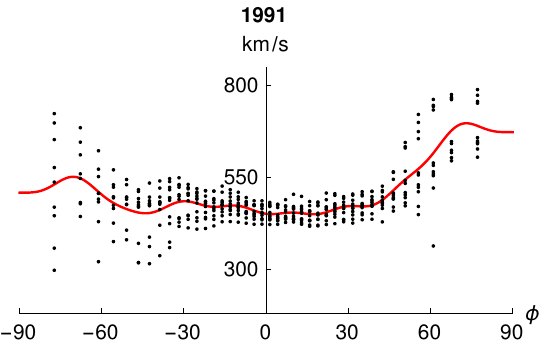}
 \includegraphics[width=0.33\columnwidth]{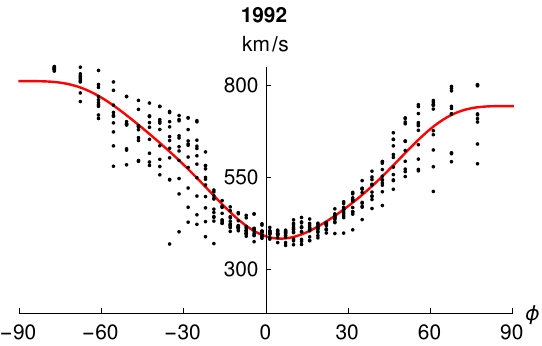}
 \includegraphics[width=0.33\columnwidth]{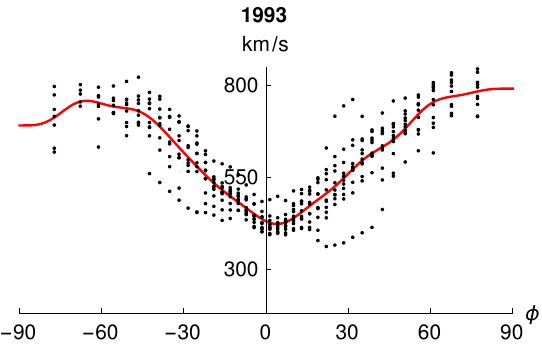}
 
 \includegraphics[width=0.33\columnwidth]{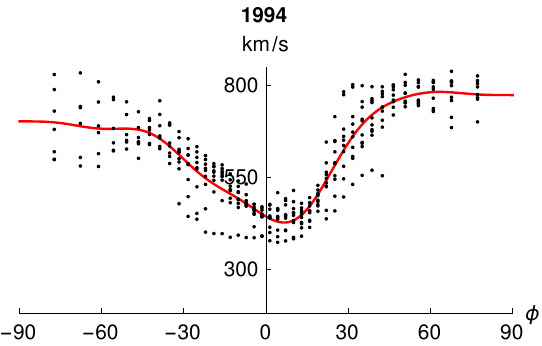}
 \includegraphics[width=0.33\columnwidth]{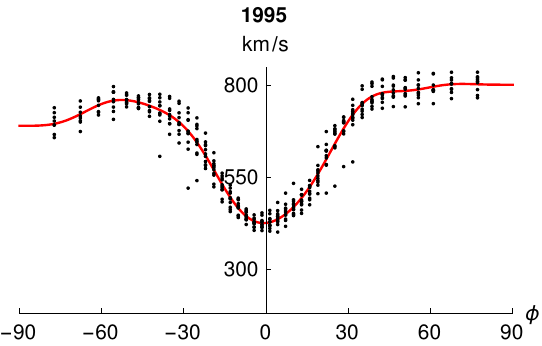}
 \includegraphics[width=0.33\columnwidth]{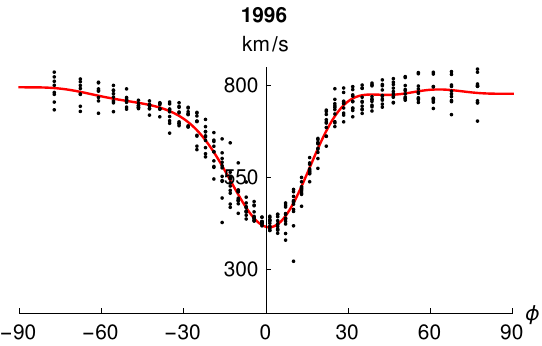}

 \includegraphics[width=0.33\columnwidth]{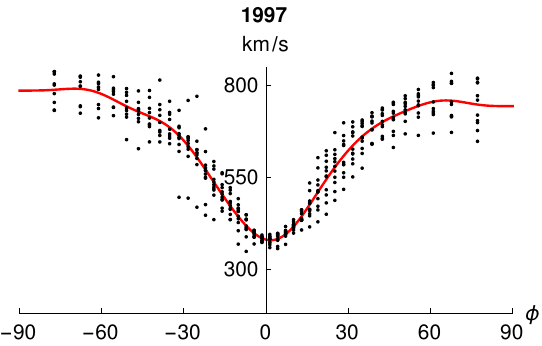}
 \includegraphics[width=0.33\columnwidth]{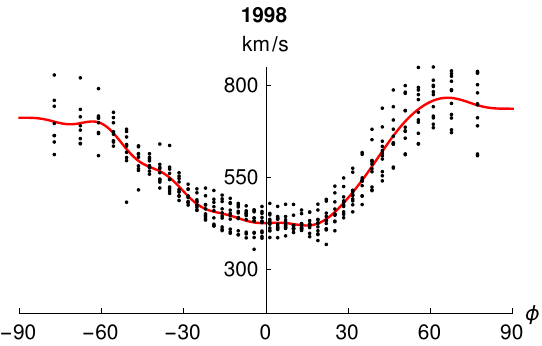}
 \includegraphics[width=0.33\columnwidth]{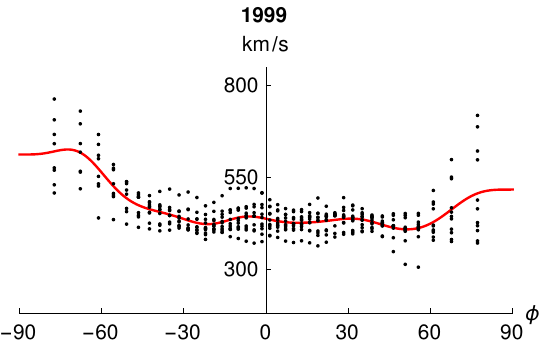}
 
 \caption{Yearly profiles of SW speed: filtered data (black dots) and the fitted model (red lines) without the OMNI adjustment for the years 1985---1999.
    }
\label{Models1}
\end{figure}

\newpage

\begin{figure}[ht]
\centering
 \includegraphics[width=0.33\columnwidth]{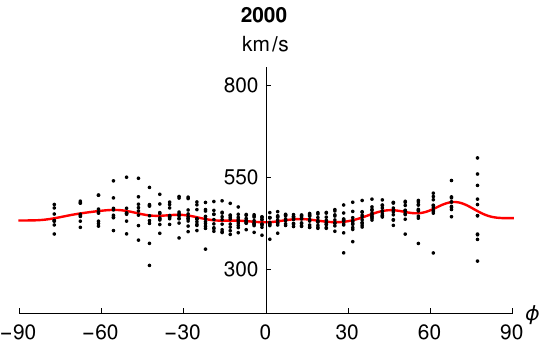}
 \includegraphics[width=0.33\columnwidth]{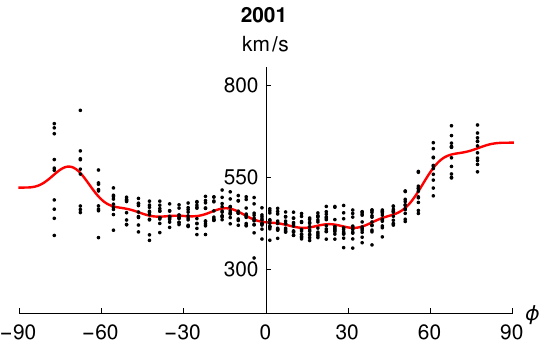}
 \includegraphics[width=0.33\columnwidth]{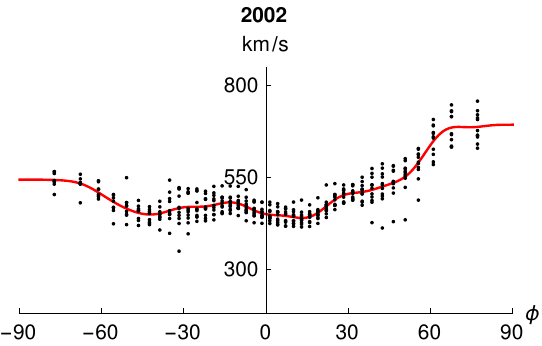}
 
 \includegraphics[width=0.33\columnwidth]{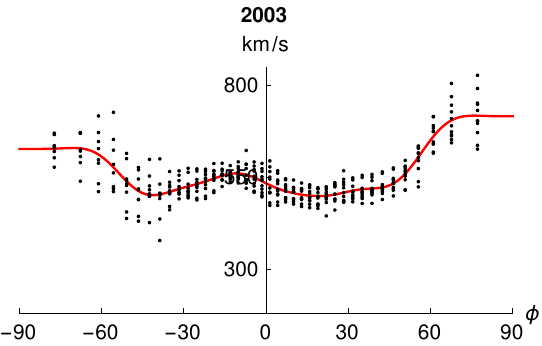}
 \includegraphics[width=0.33\columnwidth]{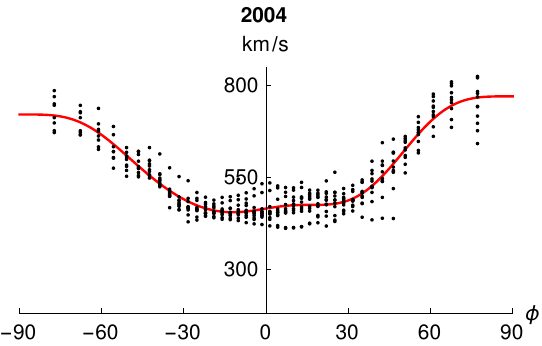}
 \includegraphics[width=0.33\columnwidth]{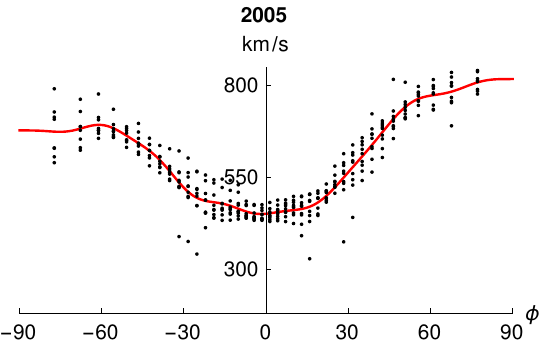}
 
 \includegraphics[width=0.33\columnwidth]{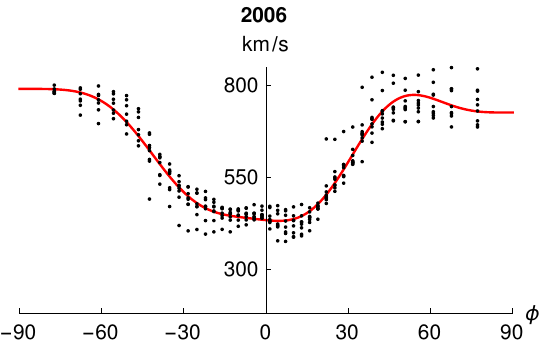}
 \includegraphics[width=0.33\columnwidth]{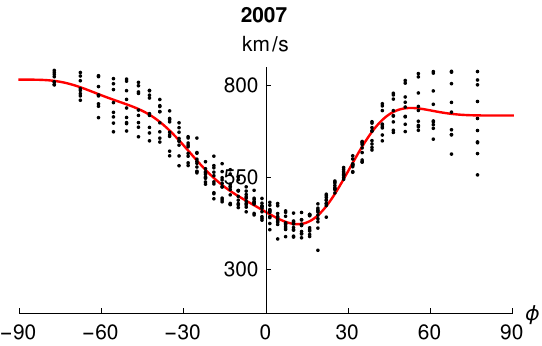}
 \includegraphics[width=0.33\columnwidth]{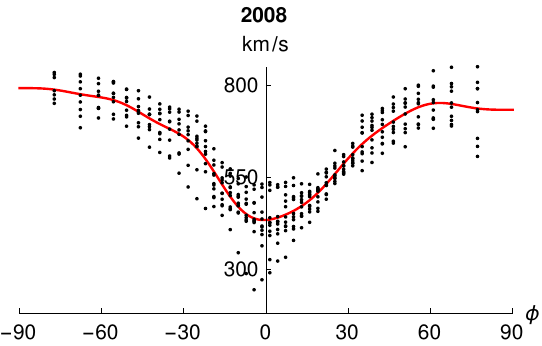}

 \includegraphics[width=0.33\columnwidth]{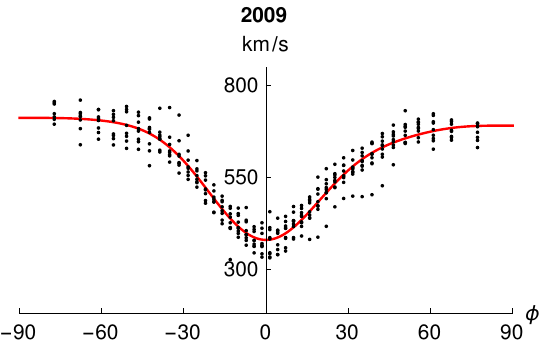}
 \includegraphics[width=0.33\columnwidth]{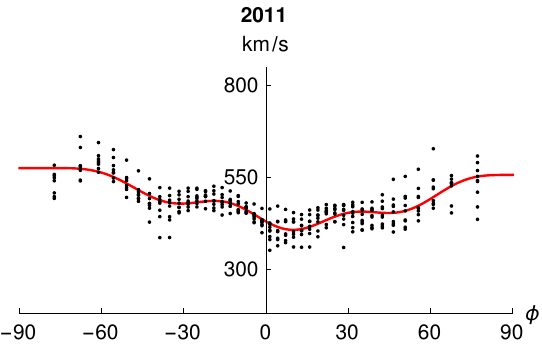}
 \includegraphics[width=0.33\columnwidth]{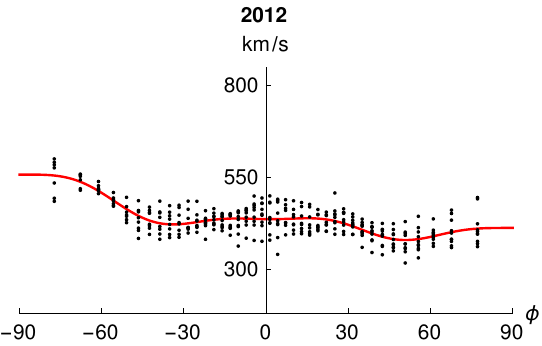}
 
 \includegraphics[width=0.33\columnwidth]{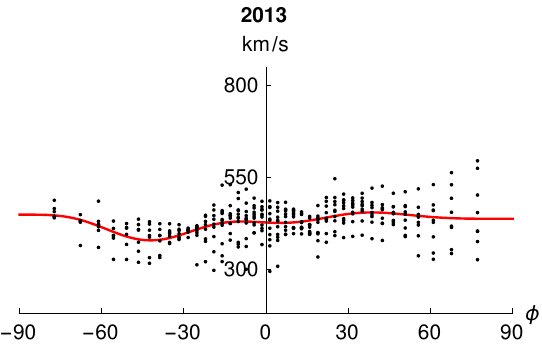}
 \includegraphics[width=0.33\columnwidth]{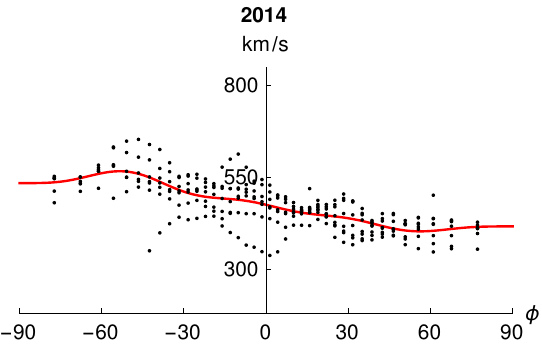}
 \includegraphics[width=0.33\columnwidth]{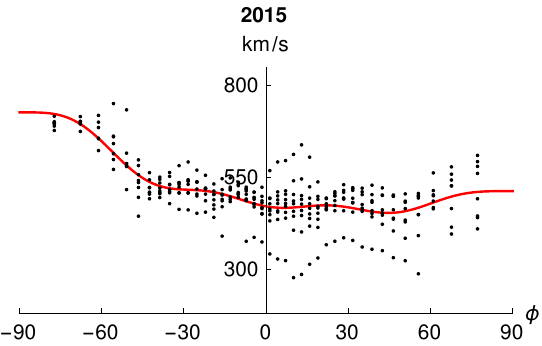}
 	\caption{Yearly profiles of SW speed: filtered data (black dots) and the fitted model (red lines) without the OMNI adjustment for the years 2000---2015. Note that 2010 is missing because of the lack of available IPS observations.}
\label{Models2}
\end{figure}

 \begin{figure}[ht]
%\centering
 \includegraphics[width=0.33\columnwidth]{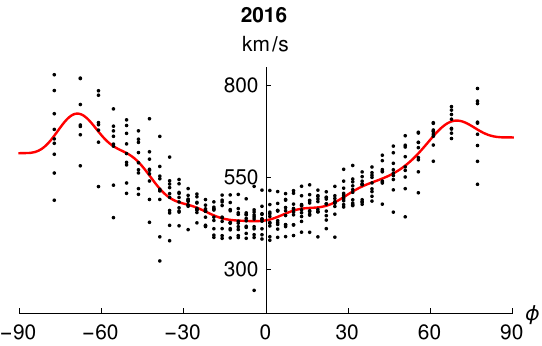}
 \includegraphics[width=0.33\columnwidth]{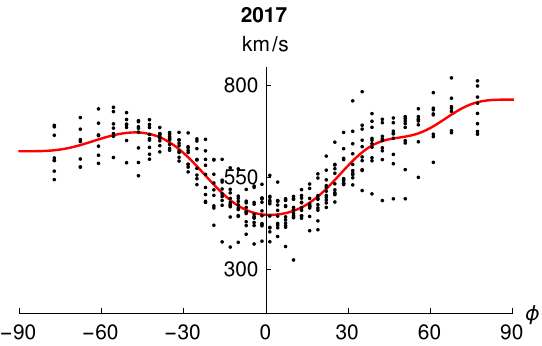}
 \includegraphics[width=0.33\columnwidth]{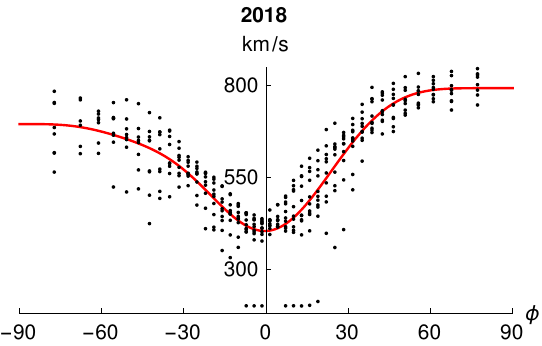}
 
\includegraphics[width=0.33\columnwidth]{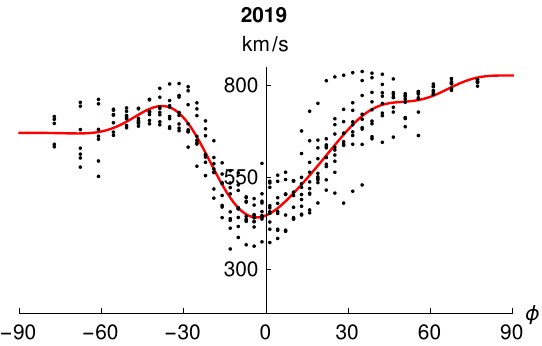}
\includegraphics[width=0.33\columnwidth]{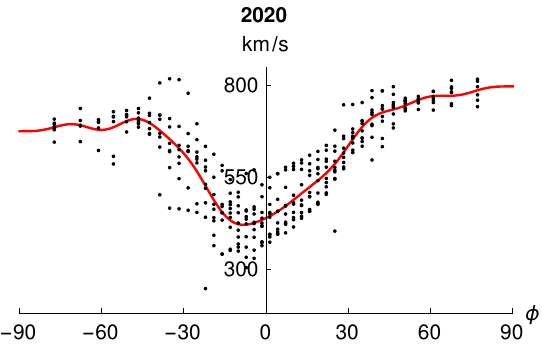}
 	\caption{Yearly profiles of SW speed: filtered data (black dots) and the fitted model (red lines) without the OMNI adjustment for the years 2016---2020.}
\label{Models3}
\end{figure}

\section{SW Speed results}
\label{sec:SWspeedresults}
At this stage, with the OMNI adjustment applied, the model can be used directly to calculate the SW speed at an arbitrary latitude during a given year. The fitted coefficients in the formula defined in Equations \ref{eq:modelLegendre} and \ref{eq:modelLegendreDeriv} for all years for which IPS measurements are available are listed in Table~\ref{tab:modelParams}. Note that application of the OMNI2 adjustment results in a decrease of the mean value of the speed over the full range of heliolatitudes. The parameter of the model that is equal to the mean value is the parameter number 1. Hence, with the adjustment magnitudes available in Table \ref{tab:omniCorrections} and the profile parameters in Table \ref{tab:modelParams}, the user can decide which version of the model to use: adjusted or not.

\begin{table}
                                                               
	\caption{Values of $Q_i$ coefficients for the best fits for all years.
    }
\tiny
    \begin{tabular}{l|r|r|r|r|r|r|r|r|r|r|r|r}
        \hline
      &1985     & 1986    & 1987 & 1988 & 1989 & 1990 & 1991 & 1992 & 1993 & 1994 & 1995 & 1996\\
	\hline
	$Q_0$   &  617.183  &  607.191  & 610.357  & 533.065   &  444.634  & 424.373  & 489.919  & 547.765  & 592.575  & 603.604  & 641.243  & 663.080 \\
	$Q_1$   &  59.8891  &  49.3345  & 48.6610  & 22.6620   &  5.63474  & 17.3161  & 39.7856  & -69.7003 & -23.8046 & 41.9977  & 26.9167  & 13.6346 \\    
	$Q_2$   &  249.114  &  175.527  & 215.166  & 263.549   &  55.2166  & 9.07071  & 95.2941  & 276.625  & 244.540  & 230.158  & 244.605  & 232.176 \\    
	$Q_3$   &  8.67587  &  -9.03749 & 21.8033  & 32.7583   &  -26.2898 & -39.6185 & 53.2847  & 45.4759  & 54.4102  & 23.1890  & 22.2639  & -15.9114\\    
	$Q_4$   &  -141.386 &  -177.342 & -151.936 & -5.37981  &  -5.81325 & -7.83491 & 53.2282  & -37.8411 & -86.4447 & -126.627 & -197.262 & -188.590\\    
	$Q_5$   &  2.86584  &  40.4999  & -12.1794 & 2.73727   &  -29.6449 & 3.12351  & -10.1236 & -5.17379 & 15.1064  & -44.8265 & 4.21816  & -13.6369\\    
	$Q_6$   &  48.5666  &  116.927  & 72.2912  & -26.3578  &  24.4394  & 4.58393  & 8.11246  & 5.76408  & 3.15663  & 22.0981  & 67.5118  & 127.523 \\    
	$Q_7$   &-          &  -23.0398 & 11.9198  & 1.68537   &  8.21345  & 8.41417  & -2.14583 &  -       & 6.25290  & 29.2850  & 13.7103  & 17.7725 \\    
	$Q_8$   &-          &  -29.3065 & -47.7847 & -10.3978  &  -14.0932 & -20.8338 & -36.8828 &  -       & -15.9056 & 18.5168  & -11.4517 & -64.7858\\    
	$Q_9$   &-         &  15.3225  & -23.5811 & -8.81113  &  25.4507  & 7.89102  & 5.48632  &  -       & -12.5051 & -27.9008 & 0.719385 & -23.4311\\    
    $Q_{10}$&-         &  -8.10741 &  11.7356 & 2.63364   &  5.77499  & -2.14633 & -25.8574 &  -       & -4.59634 & -13.3145 & -9.59446 &  -      \\    
    $Q_{11}$&-         & -         &  7.21804 & -14.4190  &  -        & 12.7055  & 4.46276  &  -       & 16.0753  &  -       & -18.3617 &  -      \\    
    $Q_{12}$&-         & -         &  -       & 4.40934   &  -        & 0.557697 & -3.16164 &  -       & -6.04774 &  -       &  -       &  -      \\    
    $Q_{13}$&-         & -         &  -       & -9.17048  &  -        & -3.46751 & -8.64356 &  -       & 4.64221  &  -       &  -       &  -      \\    
    $Q_{14}$&-         & -         &  -       & -0.726391 &  -        &  -       & -2.60835 &  -       & 11.6798  &  -       &  -       &  -      \\    
    $Q_{15}$&-         & -         &  -       & -         &  -        &  -       & -2.94092 &  -       & 2.52251  &  -       &  -       &  -      \\    
    $Q_{16}$&-         & -         &  -       & -         &  -        &  -       & -9.01438 &  -       & -        &  -       &  -       &  -      \\    
    $Q_{17}$&-         & -         &  -       & -         &  -        &  -       & -        &  -       & -        &  -       &  -       &  -      \\    
    $Q_{18}$&-         & -         &  -       & -         &  -        &  -       & -        &  -       & -        &  -       &  -       &  -      \\    
\hline

       &1997     & 1998    & 1999 & 2000 & 2001 & 2002 & 2003 & 2004 & 2005 & 2006 & 2007 & 2008\\
	\hline
	$Q_0$   & 603.846   & 541.347  & 452.036  & 442.375  & 459.633 & 495.695 & 547.449 & 538.511  & 574.040  & 579.481  & 604.499 & 612.197  \\
	$Q_1$   & -18.1253  & 21.4400  & -47.6326 & 2.54672  & 6.60358 & 55.9739 & 13.8022 & 13.1905  & 45.1956  & 29.0398  & -39.835 & -26.5515 \\    
	$Q_2$   & 278.428   & 251.031  & 70.2710  & 23.3530  & 95.5073 & 101.634 & 86.4128 & 202.317  & 236.470  & 278.574  & 255.812 & 244.066  \\    
	$Q_3$   & 8.09650   & 28.4606  & -39.6567 & 8.85651  & 46.4101 & 39.2131 & 43.5919 & 10.5880  & 21.1170  & -51.9657 & 29.1086 & 22.1796  \\    
	$Q_4$   & -152.443  & -24.5234 & 54.3667  & -8.86864 & 72.0506 & 40.4244 & 87.3678 & 49.0308  & -49.7570 & -79.5886 & -108.88 & -116.562 \\    
	$Q_5$   & -16.9948  & -33.7981 & 11.7816  & 2.02325  & -11.361 & -25.926 & -21.253 & 13.2082  & -3.36644 & -41.5602 & -65.099 & -23.2252 \\    
	$Q_6$   & 60.1243   & -56.1598 & 12.8806  & -11.5223 & 4.45963 & 14.6864 & -12.850 & -35.9874 & -40.8611 & -48.5897 & -10.199 & 33.8478  \\    
    $Q_7$   & 4.61422   & -6.52980 & 29.8524  & -2.69884 & 4.06759 & 17.9691 & 26.9754 & -16.7253 & 12.2752  & -        & 22.2975 & -8.89228 \\
    $Q_8$   & -31.1079  & 3.85576  & -2.48767 & -4.77051 & -34.215 & -30.926 & -38.915 & -        & 19.7260  & -        & 42.4266 & -19.2896 \\
	$Q_9$   &-1.04169   & 10.6474  & 0.810505 & 4.01738  & -9.3705 & -22.237 &  -16.717& -        & -2.57796 & -        & -       & 12.5585  \\
	$Q_{10}$&-2.25928  & -7.91833 & -26.7833 & -5.41395 & -14.470 & -21.492 & -17.390 & -        & 9.34537  & -        & -       & 2.10897  \\
	$Q_{11}$&5.18984   & -4.25370 & 10.5687  & -15.4981 & 17.3809 & 2.08039 & -3.0162 & -        & 4.36547  & -        & -       & -16.8816 \\
	$Q_{12}$&2.11254   & 9.67270  & -2.96951 & -9.74953 & -18.008 & 8.66556 & 10.4703 & -        & -1.76879 & -        & -       & 10.2295  \\
	$Q_{13}$&-4.96460  & -9.87217 & -8.48926 & -8.80459 & 15.3927 & 4.46332 & -3.6906 & -        & 0.525627 & -        & -       & -        \\    
	$Q_{14}$& -        & 16.0372  & 5.03811  & 2.35249  & 5.17431 & 8.67681 & 13.0607 & -        & 7.32507  & -        & -       & -        \\    
	$Q_{15}$& -        & -7.83387 & -1.14886 & 9.34817  & 8.48290 & 1.83078 & -       & -        & -15.3728 & -        & -       & -        \\    
	$Q_{16}$& -        & -7.68889 & -1.15691 & 0.581955 & 7.67939 & 0.09857 & -       & -        & -5.99806 & -        & -       & -        \\    
	$Q_{17}$& -        & -        & -        & -        & 6.39833 & 10.2356 & -       & -        & -        & -        & -       & -        \\    
    $Q_{18}$& -        & -        & -        & -        & -       & 8.79292 & -       & -        & -        & -        & -       & -        \\
\hline

      &2009     & 2011    & 2012 & 2013 & 2014 & 2015 & 2016 & 2017 & 2018 & 2019 & 2020 \\
	\hline
	$Q_0$   & 558.848   & 473.998 & 433.264 & 425.400  & 476.453  & 507.395  & 523.044  & 577.979   & 585.349  & 635.865   & 602.456  \\
	$Q_1$   & -19.8456  & -37.135 & -46.960 & 27.8035  & -84.1313 & -81.2983 & 1.04021  & 6.65843   & 61.8143  & 34.1530   & 40.6206  \\    
	$Q_2$   & 232.327   & 81.1117 & 15.3798 & 2.25655  & 7.12628  & 81.2414  & 189.722  & 177.981   & 250.515  & 200.491   & 231.793  \\    
	$Q_3$   & 5.87845   & 20.4803 & -41.341 & -25.9972 & 4.56212  & -32.3002 & -15.6351 & 51.8048   & 0.552549 & 44.5169   & 1.21636  \\    
	$Q_4$   & -124.735  & 15.0230 & 46.7080 & 17.1171  & -6.65159 & 42.2923  & 9.14089  & -101.498  & -119.077 & -168.957  & -137.055 \\    
	$Q_5$   & 8.68993   & -9.4524 & 7.12484 & -25.8637 & 23.4957  & -6.73264 & 26.5190  & 16.3340   & -21.1048 & 20.6707   & 30.4609  \\    
	$Q_6$   & 43.9744   & 13.8100 & 7.52359 & 5.71816  & -8.43182 & 14.9592  & -33.3362 & 35.0893   & 32.2694  & 98.3862   & 30.7821  \\    
    $Q_7$   & -         & 41.4056 & 11.9945 & -        &  10.1837 &  -       & 7.65180  & 6.86480   & 4.96333  & -35.1110  & -31.4381 \\
    $Q_8$   & -         & -       & -       & -        &  -       &  -       & -35.6117 & 14.7096   & -        & -0.709581 & 25.7498  \\
	$Q_9$   & -         & -       & -       & -        &  -       &  -       & 0.652647 & -0.569698 & -        & 25.3174   & 19.2606  \\
	$Q_{10}$& -        & -       & -       & -        &  -       &  -       & -24.2898 &  -        & -        & -         & -19.0662 \\
	$Q_{11}$& -        & -       & -       & -        &  -       &  -       & -6.78976 &  -        & -        & -         & -8.76906 \\
	$Q_{12}$& -        & -       & -       & -        &  -       &  -       & -19.0167 &  -        & -        & -         & 0.190140 \\
	$Q_{13}$& -        & -       & -       & -        &  -       &  -       & 19.5978  &  -        & -        & -         & 17.5816  \\    
	$Q_{14}$& -        & -       & -       & -        &  -       &  -       & 0.708739 &  -        & -        & -         & -3.04752 \\    
	$Q_{15}$& -        & -       & -       & -        &  -       &  -       & 3.19817  &  -        & -        & -         & 8.24703  \\    
	$Q_{16}$& -        & -       & -       & -        &  -       &  -       & 8.95553  &  -        & -        & -         & -        \\    
	$Q_{17}$& -        & -       & -       & -        &  -       &  -       &  -       &  -        & -        & -         & -        \\    
    $Q_{18}$& -        & -       & -       & -        &  -       &  -       &  -       &  -        & -        & -         & -        \\
                    
\end{tabular}                                                                                   
 
	\label{tab:modelParams}
\end{table}                   

\normalsize
The final step in the construction of the solar wind speed product in the WawHelIon ({\emph{Warsaw Heliospheric Ionization}}) model of the factors relevant for modeling ionization losses of neutral species in the heliosphere is a projection of the speed model on the common time and heliolatitude grid. This grid is identical to that used in the models of the heliospheric ionization factors by \citet{sokol_etal:19a, sokol_etal:20a}. This facilitates changing between ionization models in codes using the heliospheric ionization factors, such as the Warsaw Test Particle Model (WTPM) of interstellar neutral species distribution in the heliosphere \citep{tarnopolski_bzowski:09, sokol_etal:15b, sokol_etal:19b} and in the WawHelioGlow model of the heliospheric \lya{} backscatter glow \citep{kubiak_etal:21a, kubiak_etal:21b}, as well as in the calculation of survival probabilities of heliospheric ENAs between the termination shock of the solar wind and the IBEX detectors \citep{bzowski:08a, mccomas_etal:12c, mccomas_etal:17a, mccomas_etal:20a}. 

The time--heliolatitude grid in the time domain is based on a Carrington period mesh, with the nodes at the halves of the Carrington periods. In the heliolatitude domain, the grid nodes are located at multiples of 10\degr, uniformly distributed between the solar north and the south poles. The SW speed is projected on this grid. The yearly SW profiles in heliolatitude are linearly interpolated to provide speed magnitudes at halves of Carrington rotation periods and on the heliolatitude grid.

\begin{figure}[!ht]
 \centering
 \includegraphics[width=0.9\textwidth]{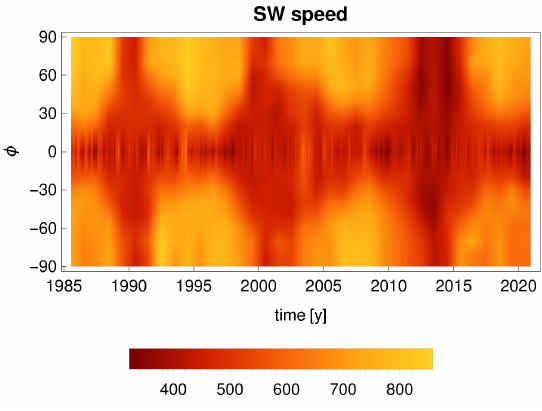}
 \caption{The evolution of the latitudinal structure of SW speed during the interval 1985--2021. The speed magnitude in \kms{} is color-coded. The model includes the OMNI2 correction and the equatorial OMNI2 insert. The time resolution is 1 Carrington rotation period, but the model values outside $\pm 20\degr$ of heliolatitude are linearly interpolated in time between the yearly profiles. 
    }
 \label{SpeedMap}
\end{figure}

We assume that the SW speed thus obtained is independent of the heliocentric distance. This assumption is well fulfilled between a few tenths of au, i.e., outside the region where the SW is being accelerated, and inside $\sim 5-10$~au. Beyond these distances, a gradual, direction-dependent slowdown of SW speed due to mass-loading from the charge-exchange between the SW and interstellar neutral H begins \citep{isenberg:86,richardson_etal:08a,nakanotani_etal:20a}.

The solar wind model we are constructing is an element of a broader Warsaw Heliospheric Ionization (WawHelIon) model, devised as a source of ionization rates of heliospheric species for interpretation of measurements. Therefore, to accommodate the much more detailed information about the solar wind parameters available at the ecliptic plane, we replace the 0\degr{} heliolatitude speeds with Carrington-averaged OMNI2 speeds, interpolated to halves of Carrington periods. The $\pm 10\degr$ nodes of the grid are filled with mean values between the equator and the latitudes $-20\degr$ and $+20\degr$, respectively. This is the last step in our SW speed modeling. The final result of the heliolatitudinal SW speed evolution is presented in Figure~\ref{SpeedMap}. The newly-constructed model will be referred to as the WawHelIon 3DSW. The fitted models of the mean yearly SW speed profiles are presented in Figures~\ref{Models1}--\ref{Models3}. The model of the SW speed, averaged in three equatorial bands identical to those used for presentation by \citet{sokol_etal:20a}, is shown in Figure \ref{vBand}. The results from this latter paper are shown along with our present results for comparison, and a discussion is provided in Section~\ref{sec:modelRobustness}. 

\section{Density}
\label{sec:density}

The SW is composed of fast and slow wind, and features a strong anticorrelation between the flow speed and the density, as clearly demonstrated by the Ulysses mission \citep{mccomas_etal:00b, mccomas_etal:08a}. Both fast and slow solar wind seem to carry the same energy flux. As shown by \citet{leChat_etal:12a}, the SW energy flux is invariable in heliolatitude within 10\% and varies weakly with time. Hence, it can be regarded as a quasi-invariant in heliolatitude and used to infer the density of the SW at an arbitrary heliolatitude provided that the speed at this latitude is known and the magnitude of the SW energy flux is known at a different latitude, e.g., in this case in the ecliptic plane. This observation has been used in the modeling of the SW structure for several years \citep{mccomas_etal:14b, sokol_etal:19a, sokol_etal:20a} and we use it in this paper.

The speed-density relation, which is linked by the SW energy flux, allows us to determine the 3D SW density model. We use the relation given by \citet{leChat_etal:12a}, also used in the previous model by \citet{sokol_etal:20a}, which has the following form:
\begin{equation}
n_p(\phi,t)=10^{-6}\bigg[m_p+\big(n_{\alpha}/n_{p}\big)(t)m_{\alpha}\bigg]^{-1}W(t)\bigg[\nu_p(\phi,t)
\bigg(0.5\nu_p^2(\phi,t)+C\bigg)\bigg]^{-1},
\label{eq:npFormula}
\end{equation}
with
\begin{equation}
W(t)=n_{p,ecl}(t)\bigg(m_p+\big(n_{\alpha}/n_{p}\big)_{ecl}(t)m_{\alpha}\bigg)\nu_{p,ecl}(t)\bigg(0.5\nu_{p,ecl}^2(t)+C\bigg),
\label{eq:Wdefinition}
\end{equation}
where $t$ is the time, $m_p$ the proton mass, $m_\alpha$ the alpha particle mass, $W$ the solar wind energy flux in $W m^{-2}$, $\nu_p$ the SW speed in \kms, $n_p$ the solar wind density in \cmc, and the constant $C=GM_\sun  R_\sun^{-1}$ is the Sun's gravity potential at the photosphere (i.e., at $R_\sun$).

\begin{figure}[!ht]
\centering
\includegraphics[width=0.45\columnwidth]{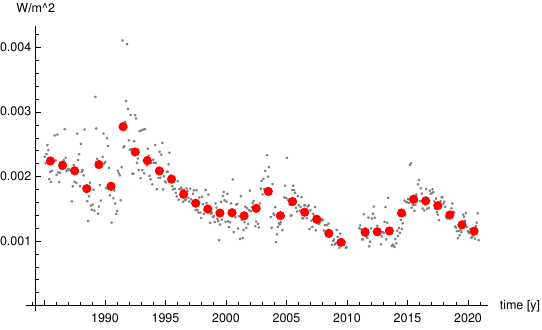}
\includegraphics[width=0.45\columnwidth]{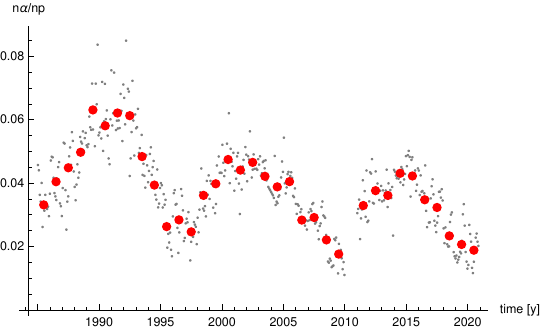}
	\caption{Left: the mean yearly values of energy flux (red points). The gray points indicate 
    this quantity averaged over individual Carrington rotations at halves of CRs. Right: the mean values of $n_{\alpha}/n_{p}$. The gray points indicate the Carrington period-averaged $n_{\alpha}/n_{p}$ values at all available halves of CRs.
	%/home/cporowski/IPSProfilesUpdate/Model2020A/Final/Model2020Density.nb
	}
\label{invariant}
\end{figure}

The SW speeds used to estimate the 3D SW density model are taken from the 3D SW speed model described above. The SW energy flux is obtained from {\emph{in situ}} measurements of the SW density, speed, and the alpha-to-proton abundance $n_{\alpha}/n_{p}$, taken from the OMNI2 time series. To find the most useful way to obtain the yearly SW energy flux magnitudes to be used for speed calculations, we first calculated  Carrington period averages of the SW energy flux in the ecliptic plane. Based on these values, we computed mean values of the energy flux for individual years using the same Carrington rotations that were used for the given year to obtain the yearly speed profiles. The yearly mean values of the energy flux thus obtained are shown in Figure~\ref{invariant}. It turned out that these values are very close to the simple yearly averages for all years in the IPS data interval.

\newpage

\begin{figure}[!ht]
  \centering
 \includegraphics[width=0.9\textwidth]{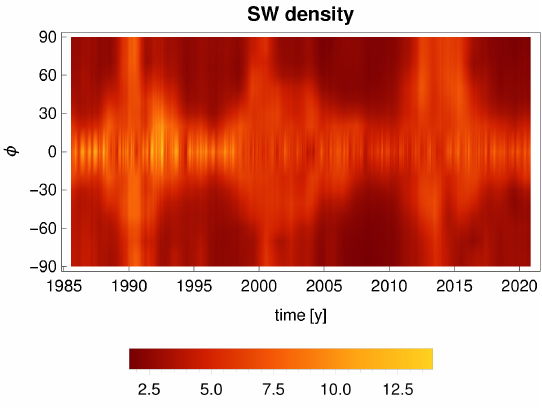}
 \caption{The model of the SW proton density evolution, presented in \cmc{} in the identical format as the SW speed model presented in Figure \ref{SpeedMap}. }
 \label{DensityMap}
\end{figure}
The yearly mean values of the SW energy flux were then inserted into Equation \ref{eq:npFormula} for all years and all heliolatitudes in the grid. This produced the yearly density profiles. Subsequently, we linearly interpolated the yearly density profiles to all CR halves similarly as we did for the speed profiles. Finally, like in the 3D wind speed model, the density values at the equator were replaced with the density values from OMNI2, while the $\pm10\degr${} bins were taken as average values from $0\degr${} and $\pm20\degr$, respectively. The resulting model of evolution of the SW density structure is presented in Figure~\ref{DensityMap}. The density model in its full resolution, organized in five heliolatitudinal bands, is presented in Figure \ref{pBand}, along with the previous model from \citet{sokol_etal:20a} for comparison. A discussion is provided in Section \ref{sec:discussionDensities}.

The density is assumed to drop with the square of heliocentric distance. This assumption is well fulfilled within a similar distance range to that for the SW speed. The density of the core SW decreases steeper than this because of the losses due to charge exchange with interstellar hydrogen atoms. However, outside $\sim 5-10$ au, in addition to the core SW population, a pickup ion population appears. It comprises former neutral H atoms that were ionized and subsequently picked up by the magnetic field frozen in the SW plasma. Pickup ions form a separate proton population in the SW. When treating the core and the pickup ion SW populations together, the total SW proton density change becomes, in fact, a small density gain relative to the $1/r^2$ relation. This is because pickup ions are created due to photoionization and ionization by electron impact. While charge exchange effectively shifts a proton from the core SW to the pickup ion population without an overall change in the proton density, the other two reactions effectively inject new protons to the pickup ion population. These ``excess'' pickup ions are former interstellar H atoms that were ionized by photoionization or electron ionization. This excess is small because the photoionization and electron impact rates are small for H atoms in comparison with the charge exchange rate \citep[cf., e.g., Figure B1 in][]{sokol_etal:20a}, which shows that the charge exchange rate is responsible for 0.7--0.8 of the total ionization rate). Hence, the assumption of a $1/r^2$ drop of the total proton density in the SW is relatively well fulfilled. However, an additional increase in the excess over $1/r^2$ relation is due to the slowdown of the SW speed due to mass-loading, discussed in Section~\ref{sec:SWspeedresults}. Detailed analysis of these subtle effects requires accounting for interaction between SW and interstellar H gas and is outside the scope of this paper.

\section{Discussion}
\label{sec:discussion}
\subsection{Is Carrington speed map filtering advantageous?}
\label{sec:filteringNeeded}

\begin{figure}[!ht]
\centering
\includegraphics[width=0.5\columnwidth]{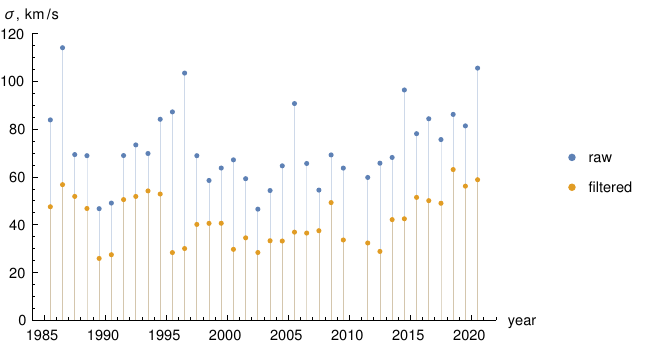}
    \caption{Comparison of the SW speed uncertainty for the models obtained from the raw and the filtered Carrington speed maps.
    %/home/cporowski/IPSProfilesUpdate/Model2020A/*Wizualizacja_i_Analiza.nb
	}
\label{ErrorComp}
\end{figure}

While filtering the data, we used the same CL for each year, to maintain the same intensity of map filtering. This is done to provide the same level of noise reduction for the entire data set, and to avoid scaling of the scatter of the residuals. Such scaling would influence the estimated uncertainty of the model. Having the same filtering level, we are able to compare the accuracy of the model with and without filtering of the map background for all individual years and to compare the effectiveness of filtering between the years. This comparison is presented in Figure~\ref{ErrorComp}.

Despite the use of a homogeneous filtering intensity, the SW speed model features  different accuracies in different years. It is probably due to different scatter of the IPS data in different years, combined with different true SW speed variations during individual years. But an improvement of the SW speed model accuracy obtained owing to map filtering is clearly visible. While the uncertainty of the model without filtering is on average $\sim70$ \kms, after filtering it is reduced to $\sim40$ \kms. An improvement of the accuracy of the model due to filtering is present for each year. For some years, the map background reduces the model inaccuracy substantially, e.g., for the year 1996, for which the uncertainty is reduced by a factor of $\sim3.5$.

\begin{figure}[t]
%\centering
 \includegraphics[width=0.3\columnwidth]{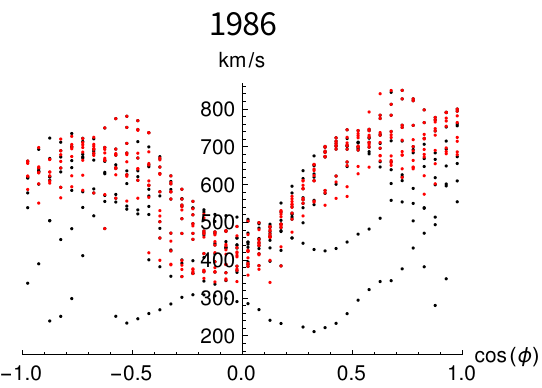}
 \includegraphics[width=0.3\columnwidth]{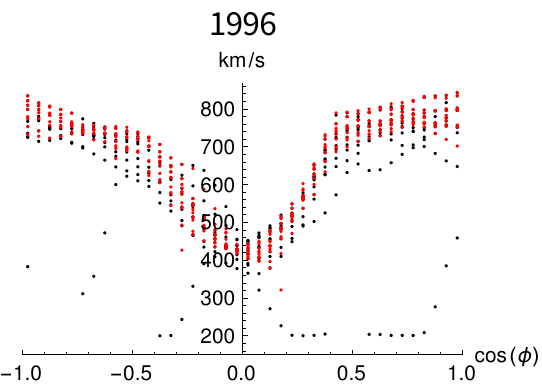}
 \includegraphics[width=0.3\columnwidth]{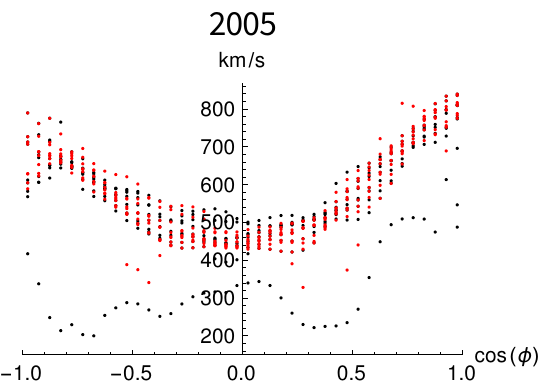}
 
 \includegraphics[width=0.3\columnwidth]{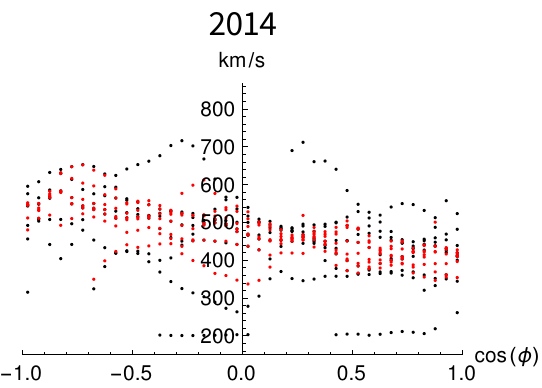}
 \includegraphics[width=0.3\columnwidth]{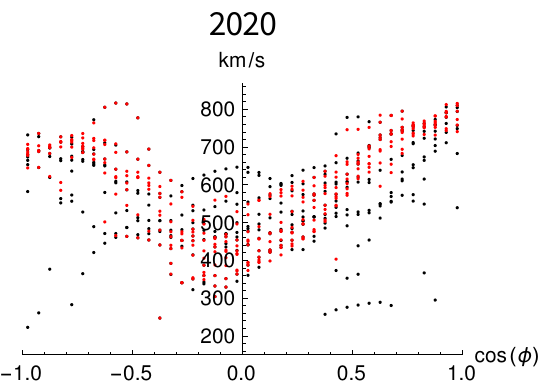}

   \caption{A review of binned speed values for the years with extreme values of the uncertainty for the raw and the filtered Carrington speed maps. The comparison shows the reasons of the extremely large uncertainty for the raw maps. 
    }
    %/home/cporowski/IPSProfilesUpdate/Model2020A/*Wizualizacja_i_Analiza.nb

\label{erroredyears}
\end{figure}

\begin{figure}[h]
	\centering
	\includegraphics[width=0.5\columnwidth]{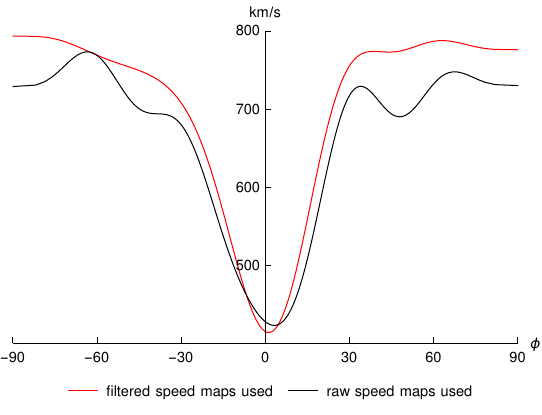}
	\caption{A comparison of a SW speed model fit for the raw and the filtered Carrington speed maps for 1996. An improvement of the fitted function due to map filtering is evident. 
    }
\label{fitcomp1996}
\end{figure}

Looking at the individual years, the years 1986, 1996, 2005, 2014 and 2020 are characterized by outstanding values of the model uncertainty, when compared to the typical uncertainty values. For these years, the uncertainty is about 100 \kms and it is significantly reduced after map filtering. A closer look at the binned speed values for these years shows the reason of such a large uncertainty for the models obtained using the raw maps (see Figure \ref{erroredyears}). In the Figure, the black points represent speed values obtained from a raw map, the red points those from a filtered map. Some of the binned profiles obtained from the raw map sharply drop to 200 \kms{} for some CRs. Such profiles are usually from the CRs at the edge of the yearly Carrington speed maps, when the coverage of data is not sufficient for a proper reconstruction of the Carrington speed maps. Our filtering procedure removed the biased points. Without filtering out the biased points, a Legendre fit to such data would be distorted, which is shown in Figure \ref{fitcomp1996}.

In this example figure, the fit for the raw map is shifted to lower speeds, and additionally a wavy behavior of the fitted model appears. The waving arises when the Legendre fitting procedure ties to model fluctuations of speed points caused by the biased speed values. The difference between the two fits reaches 80 \kms at higher heliolatitudes, which results in the biased 3D SW speed model. Also, the speeds averaged with the map background cause the profile to be wider in heliolatitude (notice that the red -- filtered -- profile is inside the black -- unfiltered profile). The filtering method is effective even for very noisy years, providing filtered speed maps, from which SW speed models are characterized by a very homogeneous final accuracy.

Summarizing this part of the discussion we point out that the filtering of the Carrington speed map background is an indispensable element to provide a reliable model based on the Legendre functions, without extensive limitations of the fitting function order. An arbitrary Legendre order limitations would be necessary to prevent introducing quasi periodic fluctuations of the model as a function of heliolatitude, but then the order would be probably too low to reproduce the salient details of the SW speed profile. Modeling of the SW speed profiles using the Legendre polynomials without map background filtering provides biased results, additionally affected by unrealistic fluctuations. Thus, the use of the filtering is advantageous not only because it removes the above mentioned drawbacks, but also because one is able to obtain objectively accurate modeling results with a reasonably low uncertainty.

Another issue is that the homogeneity of the model accuracy does not seem to be entirely random, unlike what one might expect. This is shown in Figure \ref{ErrorComp}. We may distinguish three intervals in the uncertainty plot, for which the uncertainty values have similar properties. The boundaries of the intervals seem to be identical with the periods of the $\Delta v_{IPS-OMNI}$. 
Namely, the intervals are before 1995, and after 2011. Before 1995, almost for all years the uncertainty value oscillates around 50 \kms, to drop in the next interval, which covers the Ulysses operation years.
Since 2011, the uncertainty seems to increase with time, but these variations are too low to make a statistically significant statement. Further monitoring of this uncertainty evolution will be necessary.

\subsection{The significance of OMNI2 adjustments}
\label{sec:OMICorrDiscussion}
\begin{figure}[!ht]
	\centering
	\includegraphics[width=0.5\columnwidth]{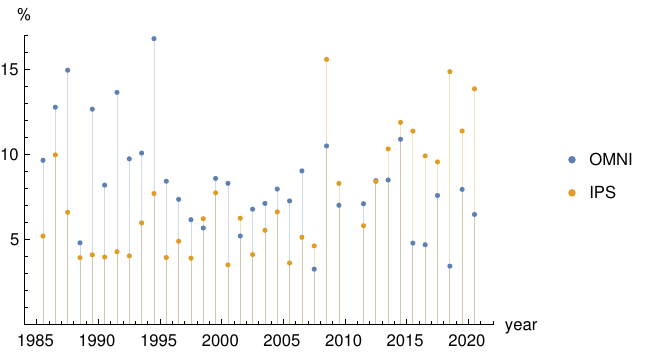}
	\caption{Relative uncertainties of the OMNI2 speed and the IPS SW speed model based on filtered Carrington maps for various years. The speed uncertainties are referred to the solar equator only.
    }
\label{IpsOmniRelErr}
\end{figure}

The OMNI2 adjustments are applied for each year individually. We check the significance of the $\Delta v_{IPS-OMNI}$ corrections compared with the effective uncertainty of IPS and OMNI2 in a given year.
First, we compare the relative uncertainties of the IPS and OMNI2 speeds at the equator to see how they differ. Such comparison is shown in Figure~\ref{IpsOmniRelErr}. The uncertainties of the IPS model are estimated from the model residuals for the equatorial area only. Since the IPS data are organized into CRs, the IPS uncertainties at the equator are mainly associated with the intrinsic variability of the average equatorial SW speeds between individual CRs in a given year. In the case of the OMNI2 data, their uncertainty is estimated from the scatter of the measurements. During estimation of the OMNI2 uncertainties, the OMNI2 data are also organized into time basis similar to the manner used for the IPS data, i.e., according to the individual CRs, which is necessary to provide the same time basis for comparison studies and to smooth variability in smaller time-scales. The individual CR numbers selected for the calculation of the uncertainty in a given year are the same for both the IPS model and the OMNI2 data, to provide the best accordance with the time coverage of the IPS data collecting periods.

Figure~\ref{IpsOmniRelErr} shows that the uncertainties of the IPS model and the OMNI2 data at the equator are of similar order. For the whole period 1985---2020, the mean relative uncertainties for the OMNI series and the model are 8\% and 7\%, respectively, with a $\sim\pm3\%$ scatter. At the beginning of the IPS data series, the relative uncertainty of the IPS model is systematically lower.
Next, a slow increase of the relative uncertainty to the level of OMNI2 appears, to finally exceed its value for the most recent years.

On the other hand, the OMNI2 uncertainty seems to drop slightly with time. It is noticeable that the relative uncertainty is the largest in 1994, when the $\Delta v_{IPS-OMNI}$ magnitude is also the largest. For the central period of the IPS data, the IPS and OMNI2 uncertainties are at the same level. An abrupt increase of the relative uncertainty for the IPS model is visible since 2008.
Despite secular changes of the uncertainties, their distributions for the IPS model and the OMNI2 data are similar according to Pearson's $\chi^2$ test and are not correlated. So the uncertainties may be treated as independent.

\begin{figure}[!ht]
\centering
\includegraphics[width=0.7\columnwidth]{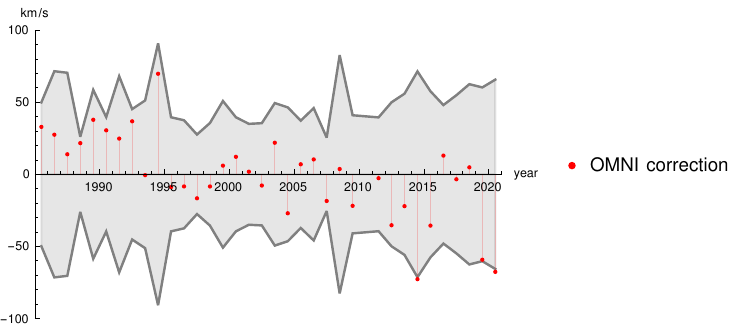}
    \caption{The values of the OMNI2 adjustments $-\Delta v_{IPS-OMNI}$ (red points) compared with the effective 	uncertainty of the IPS and OMNI2 (shaded area).}
\label{fig:IPScorr}
\end{figure}

Now we study the significance of the $-\Delta v_{IPS-OMNI}$ adjustments relative to the effective uncertainty of the IPS and the OMNI2 speed components used to calculate the adjustments (see Fig. \ref{fig:IPScorr}). The comparison shows that the significant $-\Delta v_{IPS-OMNI}$ adjustments are obtained for the following years: 2014, 2019 and 2020, i.e. for 9\% of all processed years (see Figure
\ref{fig:IPScorr}).

Summarizing this part, we see that the scatter in the OMNI2 data is similar to that of the SW modeled speed, and for both data sets the relative errors are not correlated. Also, if a time trend exists in the IPS data, which at the moment is in question, it lies within the statistical accuracy of the model, so it should not introduce any systematic errors to the results. Taking the above into account, and having in mind that for the equator the yearly averaged OMNI2 and IPS SW speed models have the same statistical properties, and the adjustments are not statistically significant for 91\% of the years in comparison with the model accuracy, the shifting of the helioatitude profiles of the model by $-\Delta v_{IPS-OMNI}$ seems appropriate. Statistical properties of the OMNI2 and the SW speed model uncertainty confirm that no significant systematic errors will be introduced this way. An additional argument in favor of application of the adjustments is that the OMNI2 speeds are derived from solar wind speeds measured at various spacecraft and cross-calibrated, so systematic differences resulting from application of different instrument and data analysis techniques are largely eliminated, and application of the adjustments to the IPS-derived speed profiles provides a good correspondence of the IPS SW 3D speed model with the OMNI2 data in the ecliptic plane.

Adoption of a different approach, i.e. adopting the hypothesis that a time trend in $\Delta v_{IPS-OMNI}$ indeed exists, would cause nasty practical problems. The IPS observations suitable for our analysis arrive at a yearly cadence. An analytic formula would need to be fitted to the $\Delta v_{IPS-OMNI}(t)$ time series. With each new year of data, the entire set of corrections to the yearly SW speed profiles would need to be applied, which would result in a new version of the entire data set starting from 1985. Given the fact that none of the simple analytic models we have tested shows a clear advantage, we decided to apply individual yearly corrections. 

\subsection{The robustness of SW speed modeling}
\label{sec:modelRobustness}

\begin{figure}[!ht]
\centering
\includegraphics[width=0.9\columnwidth]{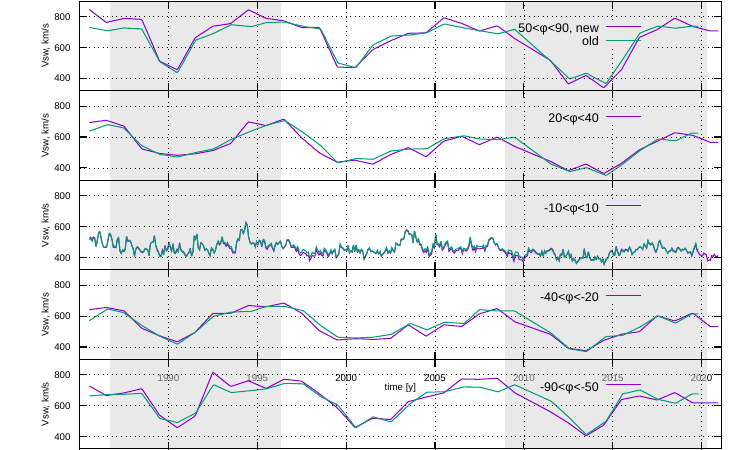}
    \caption{Our model of the SW speed, based on filtered Carrington speed maps for the years 1985.5 to 2021, shown in magenta. The shaded regions encompass Solar Cycles (SC) 22 and 24. For comparison, the model by \citet{sokol_etal:20a}, which uses the previous version of IPS Carrington speed maps and a different approximating function for the latitudinal speed profiles is presented in green.
    %/home/cporowski/IPSProfilesUpdate/Model2020A/Final/plotPubl.inp
	}
\label{vBand}
\end{figure}

The overall characteristics of the modeled SW speed is very similar to the previous model, which is shown in Fig. \ref{vBand}, but some differences are evident. Generally, while the WawHelIon 3DSW model tends to provide higher speed values during the solar minima at higher heliolatitudes, during the solar maxima the models are very consistent. This is seen at the beginning of the data set, as well as around 1994, when an exceptionally high difference between the OMNI2 speeds and those from IPS observations are present. In 1994, a comparison between the new and the previous model shows a kind of new feature for this year in the form of a deficit in the previous model, distinguishable especially in the north hemisphere. This feature is undoubtedly linked with a large difference between the OMNI2 and the IPS data for 1994, which was not present before, and should be monitored in future releases of the IPS data. Also, around 2008 and 2018 the WawHelIon 3DSW model provides systematically higher speeds, but for these years the difference in speed between the two models is smaller than around year 1994.

Looking at the dynamics of the SW speed, for the beginning of Solar Cycle (SC) 24 the WawHelIon 3DSW model provides an earlier decline of the SW speed. It is noticeable that this appears around year 2010. The difference in the dynamics is much larger than at the beginning of SC23 (between 1995 and 2000), when the decline started almost simultaneously in the previous and current models. Despite the earlier decline, the SW speed becomes consistent between the two models during the solar maximum.

\begin{figure}[!ht]
\centering
 \includegraphics[width=0.32\columnwidth]{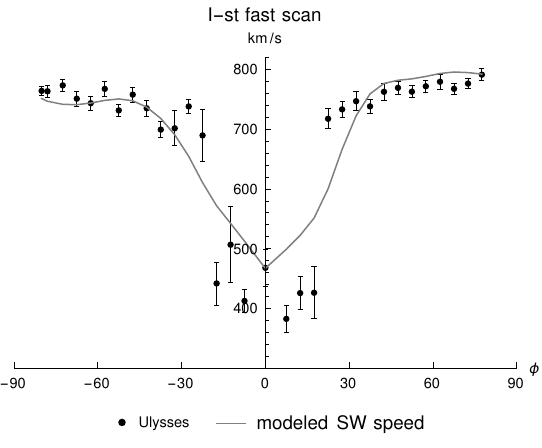}
 \includegraphics[width=0.32\columnwidth]{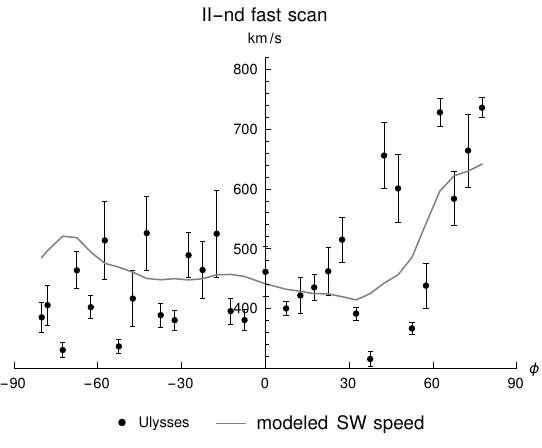}
 \includegraphics[width=0.32\columnwidth]{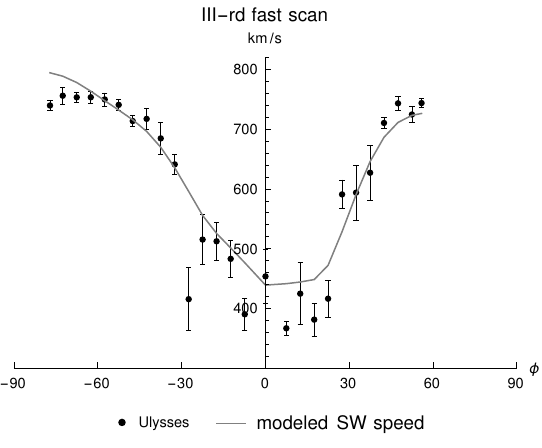}
  \caption{Comparison of 3D SW speed model with \it in situ \rm measurements.
    Error bars for OMNI2 data indicate 1$\sigma$ uncertainty taken from data scatter.
    %/home/cporowski/IPSProfilesUpdate/Model2020A/Final/UlyComp.nb
	}
\label{UlyComp}
\end{figure}

\begin{figure}[!ht]
\centering
\includegraphics[width=0.9\columnwidth]{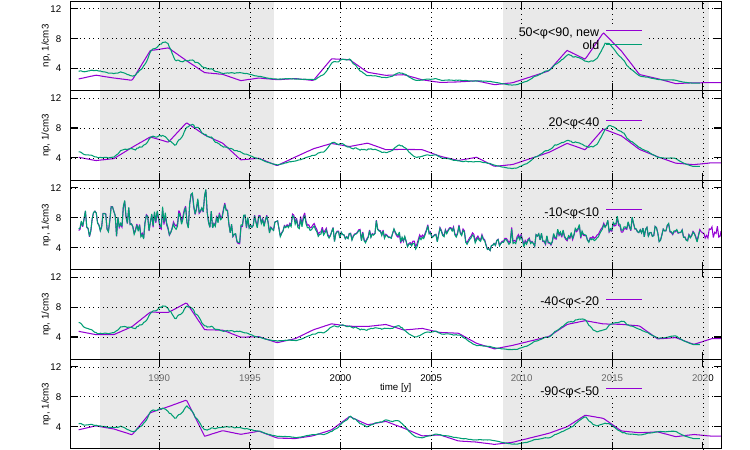}
    \caption{
	    The model of the SW density based on the filtered Carrington speed maps and the SW energy flux invariant for the years 1985.5 to 2021. The shaded regions encompass SC 22 and 24. The magenta line is our present model, the green lines represent the model by \citet{sokol_etal:20a}, shown here for comparison. 
    %/home/cporowski/IPSProfilesUpdate/Model2020A/Final/plotPubl.inp
	}
\label{pBand}
\end{figure}

We also compare the 3D IPS SW speed model with the \it in situ \rm measurements from fast Ulysses scans.
The Ulysses data are averaged in 5\degr equatorial bands. The uncertainties were derived from the scatter of the data within the bins. The comparison is presented in Figure \ref{UlyComp}. A general good agreement with \it in situ \rm measurements is seen during the first and the third fast
scans. During second fast scan the Ulysses data show a large scatter due to physical changes, but the overall characteristics seems to agree.
\color{black}

\subsection{Comparison of the densities}
\label{sec:discussionDensities}

We point out an important difference between our approach to the calculation of the SW density and that used by \citet{sokol_etal:20a}. While we derive the density model for a given year from the corresponding mean yearly speed profile and the latitudinally invariant mean yearly solar wind energy flux, \citet{sokol_etal:20a} use running yearly averages separately for each Carrington rotation  of the invariant to calculate the Carrington period density profiles. The Carrington averages of the SW energy flux feature a larger scatter than the yearly averages, and these fluctuations are propagated into all heliolatitudes. As a result, the speed model by \citet{sokol_etal:20a} features smooth variations of speed at heliolatitudes outside the equatorial band, in the case of the density, these variations are much more rapid. We are not convinced these variations are realistic since we are not sure if the invariance of the SW energy flux is maintained on such a short time scale as one solar rotation period. Therefore, we decided to use the approach presented above. After all, the existence of this invariant was based on pole-to-pole observations from Ulysses, as pointed out by \citet{leChat_etal:12a}, and the time of Ulysses fast latitude scan between the poles was approximately 6 Carrington rotations \citep[see, e.g., Figure 1 in][]{bzowski_etal:14a}. The invariance at shorter time scales may exist but to our knowledge, has not been confirmed.

Nevertheless, a comparison between the new and old models shows general similarities. Typical density variations correlated with the solar maxima and minima are present. Also, the north-south asymmetry in the density, as in the previous model, is present. Not all two-peak structures in time, present in the previous model, are confirmed by the new model. This may be due to the different approach to the calculation of the density that we used. Our model confirms the presence of the two-peak structure at the northern hemisphere in the middle latitudes for SC 22 and 24, and at high latitudes only for SC24. At the southern hemisphere, only in middle-high latitudes the two-peak structure appears during SC23.

A comparison of the absolute values of the densities between the two models shows that systematically higher values are obtained only for the highest latitudes in both hemispheres in SC24. The density values at the peak during SC24 at the highest latitudes in the northern hemisphere, returned by the WTPM 3DSW model, are the highest in the entire interval, reaching above 8 cm$^{-3}$. Also, in the southern hemisphere the WTPM 3DSW model provides high and more stable density values, while the previous model provides a lower density between the two-peak structures, i.e., around 1992 and 2014. It is likely that the lower value in the two-peak structures may be linked with the different approach to averaging and interpolating of the models.

A portion of the differences between the densities from the two models is due to the differences in the speed. They are propagated to the densities by the relation in Equation~\ref{eq:npFormula}. Another portion of the differences comes up because of the different approach to the application of the invariant. 

\subsection{Concluding discussion}
\label{sec:concludingDisc}
The agreement between WTPM 3DSW and the model by \citet{sokol_etal:20a} in both speed and density are quite good, even though the original IPS data had been modified between these two releases of the model, as well as data filtering method and the analytic form of the fitting functions were changed. This illustrates the maturity and robustness of the line of modeling the 3D structure of the solar wind based on IPS CAT analysis and the approximations used in the papers by \citet{sokol_etal:13a, bzowski_etal:13a, sokol_etal:15d, sokol_etal:19a}, and the present paper.

Analysis of the helioglow aimed at retrieval of the latitudinal structure of the solar wind, performed based on observations of the helioglow from the SWAN experiment onboard the SOHO mission \citep{bertaux_etal:95} and presented in a series of papers by \citet{lallement_etal:10b, katushkina_etal:13a, katushkina_etal:19a, koutroumpa_etal:19a} suggested that the solar wind flux may feature maxima at mid-latitudes. Our analysis of the SW speed structure, presented in figures \ref{Models1}--\ref{Models3}, shows that while indeed, local minima and maxima of speed exist in different years, usually they are not pronounced, they are not regular and they occur at different latitudes. This may be an important finding because, unlike the previous model, WawHelIon 3DSW is able to reproduce any profile that is supported by the data. Hence, should the features suggested by these authors be real, they would be due to the evolution of the density with heliolatitude, but then very likely the reality of the SW energy flux invariance with heliolatitude would be challenged. This question certainly requires further study.

\section{Summary and conclusion}
\label{sec:conclusions}

We present a new model of evolution of the latitudinal structure of the SW speed and density, calculated from a new IPS-derived data set of SW speed: the WawHelIon 3DSW model. The input data cover a newly-reprocessed entire set of available IPS observations subjected to CAT analysis, with an important addition of variable coefficient $\alpha$, connecting the magnitude of the fluctuations of the SW electron density $\Delta n_e$ with the SW speed $V$: $\Delta n_e \sim V_{SW}^\alpha$. The model covers a time interval from 1985 until the end of 2020.

We developed and applied a new method of filtering of the IPS-derived Carrington maps of solar wind speed against outlying values. This method acknowledges that some of the data points in the Carrington SW speed maps may be biased. It identifies the suspect outlier points based on the ESD method without arbitrary assumptions and returns a filtered data set, suitable for approximation by an analytic function. We calculated yearly-averaged heliolatitudinal profiles of SW speed by fitting the filtered yearly data with an analytic formula based on Legendre polynomials of an order that can vary between the years, and with an additional condition of a null first derivative over the poles. The order of the formula for individual years was determined based on statistical analysis of the residuals. The formula is flexible enough to be able to reproduce even small-scale variations, should they be identified, and does not overfit the data. 

We performed an extensive comparison of the predictions of the model with in-situ SW time series of the SW speed from the OMNI2 collection and requested that the two data sets agree at the solar equatorial plane. To that end, we studied differences between the OMNI2 and our model predictions and derived necessary corrections based on a meticulous statistical analysis. 

We calculated yearly heliolatitudinal profiles of the SW density using yearly averaged corrected SW speed profiles assuming that the SW energy flux is latitudinally invariant on a time scale of one year. Finally, we projected the SW speed and density models on a regular heliolatitudinal grid, with time nodes at halves of individual Carrington rotation periods. The equatorial band was replaced with appropriate Carrington period averages of the SW speed and density from the OMNI2 time series.

The newly developed model can be used in its tabular form, as a grid of densities and speeds tabulated in heliolatitude and time, or in its analytic form for individual years. 

The results of the WTPM 3DSW model and that of \citet{sokol_etal:20a} generally agree, even though the latter one used the former version of the input Carrington maps of SW speed, a different method of data filtering, and a different analytic approximating function. Also different was the assumptions made on the invariance of the solar wind energy flux. The WTPM 3DSW model suggests higher SW speed values during solar minima and a faster decline of SW speed at the beginning of SC24. The higher speed values at solar minima may be caused by the elimination of a bias in Carrington maps caused by outlier data in all CRs during all years, which was not fully addressed in \citet{sokol_etal:20a}.

The SW densities provided by the WTPM 3DSW model are also in general agreement with that by \citet{sokol_etal:20a}, but a part of fine features seen in the model by \citet{sokol_etal:20a} as two-peak structures disappeared in our model. This is likely due to the different use of the solar energy flux invariant. Since it is currently unknown whether or not the invariance of the solar wind energy flux holds on a time scale of one solar rotation, as assumed by \citet{sokol_etal:20a}, or a time scale of 1 year, as assumed in the present paper, this discrepancy provides an estimate of the uncertainty of the density of SW at high latitudes. The general agreement between the two models suggest the robustness of data analysis and methodology adopted at all phases of research in the line of papers using the IPS observations and analytic function fitting that gives the time evolution of the SW structure.

The WawHelIon 3DSW model of the evolution of SW parameters is well suitable for use in global heliospheric studies, as well as analysis of heliospheric in-situ measurements. In particular, it can be used to compute attenuation of energetic neutral atoms between their origin in the outer heliosheath and detectors located deep in the inner heliosphere, like IBEX and the planned IMAP missions at 1 au.  

\begin{acknowledgments}
{\emph{Acknowledgments}}. The authors gratefully acknowledge helpful discussions with Marzena A. Kubiak. The OMNI data were obtained from the GSFC/SPDF OMNIWeb interface at https://omniweb.gsfc.nasa.gov. This study was supported by Polish National Science Center grant 2019/35/B/ST9/01241.
\end{acknowledgments}

\bibliographystyle{aasjournal}
\bibliography{iplbib}

\end{document}